# Partitioning, Indexing and Querying Spatial Data on Cloud

by

Afsin Akdogan

A Dissertation Presented to the
FACULTY OF THE USC GRADUATE SCHOOL
UNIVERSITY OF SOUTHERN CALIFORNIA
In Partial Fulfillment of the
Requirements for the Degree
DOCTOR OF PHILOSOPHY
(COMPUTER SCIENCE)

December 2015

# Acknowledgments

I dedicate my dissertation work to my loving family for motivating and supporting me throughout my life. I would like to thank my advisor, Professor Cyrus Shahabi. He has been a true mentor during my PhD; from whom, I have learned innumerable thing. In addition, I would also like to thank my PhD guidance committee members for their valuable feedback and being a part of my story. Finally, many thanks to my friends in Los Angeles who have always been by my side to cheer me up.

# Table of Contents

# List of Tables

# List of Figures

## Abstract

The number of mobile devices (e.g., smartphones, wearable technologies) is rapidly growing. In line with this trend, a massive amount of spatial data is being collected since these devices allow users to geo-tag user-generated content. Clearly, a scalable computing infrastructure is needed to manage such large datasets. Meanwhile, Cloud Computing service providers (e.g., Amazon, Google, and Microsoft) allow users to lease computing resources. However, most of the existing spatial indexing techniques are designed for the centralized paradigm which is limited to the capabilities of a single sever. To address the scalability shortcomings of existing approaches, we provide a study that focus on generating a distributed spatial index structure that not only scales out to multiple servers but also scales up since it fully exploits the multi-core CPUs available on each server using Voronoi diagram as the partitioning and indexing technique which we also use to process spatial queries effectively. More specifically, since the data objects continuously move and issue position updates to the index structure, we collect the latest positions of objects and periodically generate a read-only index to eliminate costly distributed updates. Our approach scales near-linearly in index construction and query processing, and can efficiently construct an index for millions of objects within a few seconds.

In addition to scalability and efficiency, we also aim to maximize the server utilization that can support the same workload with less number of servers. Server utilization is a crucial point while using Cloud Computing because users are charged based on the total amount of time they reserve each server, with no consideration of utilization. Therefore, it is essential for users to fully utilize all servers to reduce the cost. Clearly, one key factor that impacts server utilization is the partitioning method especially in data-driven location-based services. This is because if the data partitions are not accessed, the servers storing them remain idle but the user is still charged. To this extent, we also consider the access patterns (workloads) of our Voronoi-partitions and study a cost-effective partitioning method that aims to reduce the server cost.

# Chapter 1: Introduction

## 1.1 Motivation

The number of mobile devices (e.g., smartphones, wearable technologies) is rapidly growing. In line with this trend, numerous location-based services/applications have emerged that use the moving object datasets generated by such mobile devices e.g., mobile phone social networking, spatial crowdsourcing, scientific simulations, and ride-sharing applications such as Uber. These applications need to handle a tremendous number of spatial objects that are continuously moving and executing spatial queries to explore their surroundings. Obviously, as required by these applications, processing a large volume of spatial queries (e.g., range, *k* nearest neighbor) on these massive datasets in the presence of high update workload presents a major *scalability* problem.

Meanwhile, Cloud Computing service providers such as Amazon and Microsoft allow users to lease computing resources that can scale instantaneously. However, most of the existing approaches, which are devised to handle such update-heavy workloads, are primarily designed for the **centralized paradigm** which is limited to the capabilities of a single server. For example, specialized moving-object index structures have been proposed which maintain location updates incrementally and process only a fraction of the updates using position approximation techniques. However, all of these approaches suffer from limited scalability due to their single-server design. Alternative to this incremental model, a few studies adapt throwaway model, where a static index is periodically created from scratch and costly updates are not processed in-place. In this model, queries are forwarded to the current index, while the latest positions of the data objects are collected in a buffer (e.g., hash table). Subsequently, a fresh index is effectively created from scratch with bulk-loading and swapped with the current index. It has been shown that periodically rebuilding a static index significantly outperforms specialized moving-object indices in all but the most extreme cases. However, throwaway indices suffer from scalability due to their single-server design too.

The need for our research arises from the fact that as the applications gain popularity they all face a major scalability problem and we still lack the proper index structures that can scale out to multiple cloud servers and utilize emerging computing infrastructures. In our first study, we observed that a distributed throwaway index, which can **scale out** near-linearly as the number of servers increases, can efficiently be constructed within seconds for millions of data objects using Voronoi diagrams (VD). Indeed, Voronoi diagram can be used as 1) partitioning technique to distribute the data objects across multiple servers and 2) underlying index structure to process queries at each server efficiently. In addition to scaling out to multiple servers, it is also crucial to utilize multi-core CPUs to **scale up** as chip vendors such as Intel have recently shifted to multi-core architecture rather than increasing processor speeds, which is projected to be the trend in the next 15 years. Voronoi Diagram is a perfect fit to utilize the multi-core CPUs at each server as well since each Voronoi cell can be built autonomously. Therefore, we are interested in to build a Voronoi diagram in a distributed fashion for continuously moving objects using multiple servers and multicore CPUs.

Even though Cloud Computing provides flexible resources, the total cost increases with the total number of servers even if they are not fully utilized. This is because the largest cloud computing service providers (e.g., Amazon EC2, Microsoft Azure, and Google Compute) adapt a **time-based** pricing model and charge users based on the total amount of time for each server with no consideration for the percentage of utilization, as long as the servers are running. We would like to note that there are several cloud data stores where users pay only for the resources they actually consume, unlike the time-based pricing model. For example, Amazon SimpleDB and DynamoDB charge users for each read and write operation. However, these services are not appropriate to manage multi-dimensional spatial data because they mostly support simple key-value based put-get operations with either no or very limited spatial query support. Whereas, with spatial data, it is expected and crucial to provide advanced querying features (e.g., range, k nearest neighbor, skyline). Thus, for location-based services it is crucial to fully exploit each and every server as much as possible. One key factor that impacts server utilization is the partitioning method especially in data-driven location-based services. This is because if the data partitions are not accessed, the servers storing them remain idle but the user is still charged. Whereas, existing spatial data partitioning techniques aim to 1) cluster spatially close data objects in the same node to minimize disk I/O and 2) create equi-sized partitions. On the contrary, the objective is different for cloud given the current pricing models. Therefore, we aim to devise a **cost-efficient** partitioning method for spatial data where an increase in the servers' utilizations yields less number of servers to support the same workload, thus saving cost. To achieve this goal, we aim to spatially partition data objects using Voronoi diagram and integrate access patterns (workload) associated with the Voronoi partitions (e.g., how many times a partition is accessed at a particular time) to maximize serer utilization.

## 1.2 Research Significance and Applications

The significance of our studies lie in our proposed solutions to utilize the emerging computing infrastructures more efficiently, tackle the real-world problems that were inadequately addressed and creating new opportunities for future studies.

The distributed throwaway index structure opens up new opportunities for database researchers to study a paradigm shift in indexing. Computing systems have undergone a fundamental transformation from the relatively isolated single-processor devices to networked devices with 100s of CPUs via the cloud. Therefore, it is now possible to devise parallel index construction algorithms that may result in discontinuation of traditional incremental index maintenance for moving object datasets. In addition, such parallel scalable algorithms, which can scale out to multiple servers and scale up to multi-core CPUs, help the worldwide community of researchers to quickly solve important problems regarding geospatial data. For instance, neuroscientists model the neural system as a set of spatial objects and simulate how the neurons move over time, where every object i) changes its location at every simulation time step and ii) runs spatial queries to discover other objects around them. Obviously, as required by these applications, processing a large volume of queries on these massive datasets in the presence of high update workload presents a major scalability problem and results in unreasonable execution times (e.g., several hours). With parallel scalable algorithms, the same problems can efficiently be solved within few seconds. In addition, the potential impact of our proposed research goes beyond computer science and delivers significant environmental benefits. In particular, the outcome our research can facilitate reducing energy consumption and carbon emission by employing an elastic solution which uses a cluster of servers that can grow and shrink very quickly in seconds depending on the demand.

The research significance of cost-efficient partitioning is centered on the inclusion of **workloads** associated with data partitions as another and new way of partitioning large dataset across multiple servers. The main goal of such inclusion is maximizing server utilization, and hence reduce the server cost along with the number of servers. Existing partitioning approaches adapt schema-based strategy aiming to generate equi-sized partitions. As a practical example, MongoDB (a popular NoSQL database) only balances the amount of data at each server without considering whether the data is accessed or not. As for applications, cost-efficient partitioning can be used by global location-based services such as Yep -a popular online urban guide with over 100 million monthly users. Users typically use Yelp during daytime in their time zone to find local businesses around them and may not use this service as frequently during nighttime. When a conventional spatial partitioning technique is used to distribute such a geospatial dataset across multiple cloud servers, this correlation causes a remarkable amount of underutilization of the servers. Suppose we distribute Yelp's data across 100 servers using Voronoi diagram without considering workloads, where each server stores a disjoint subset of the data. Suppose as a result of this partitioning 50 servers are allocated to store the data associated with USA and the rest for Europe. In such a case, either batch of 50 servers will be idle almost half of the day due to infrequent accesses of the data at midnight in their time zone.

## 1.3 Preliminaries

**Voronoi Diagram:** A Voronoi diagram decomposes a space into disjoint polygons (cells) based on the set of generators (i.e., data points). Given a set of generators S in the Euclidean space, Voronoi diagram associates all locations in the plane to their closest generator. Each generator s has a Voronoi cell consisting of all points closer to s than to any other site. Hence, the nearest neighbor of any query point inside a Voronoi cell is the generator of that cell.

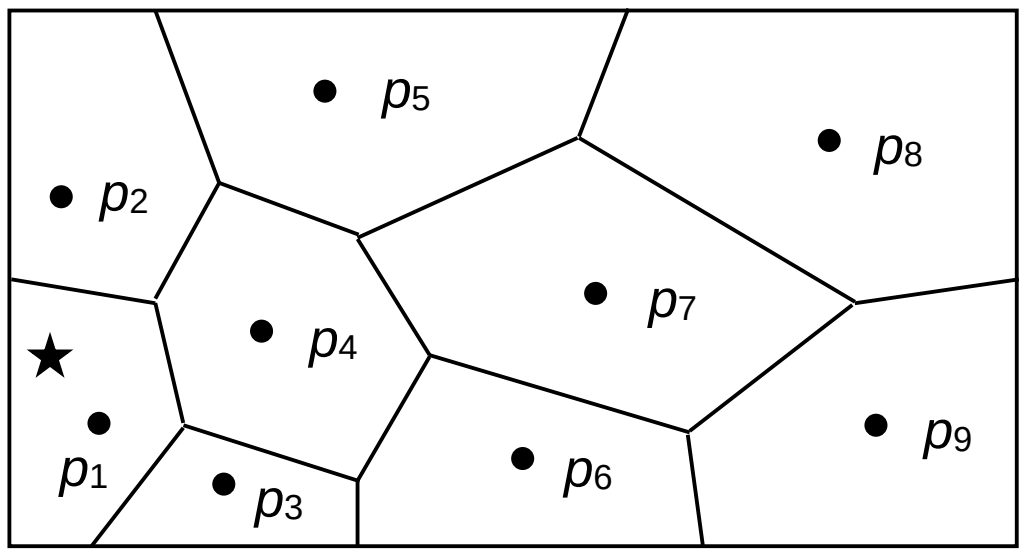

Figure 1.1: A sample Voronoi Diagram

*DEFINITION 1. Voronoi Cell*

Given a set of generators $P = \{p_1, p_2,\ldots,p_n\}$ where $2 < n < \infty$ and $p_i \neq p_j$ for $i \neq j$, $i, j = 1;\ldots, n$, the Voronoi Cell of $p_i$ is $VC(p_i) = \{p \mid d(p, p_i) \leq d(p, p_j)\}$ for $i \neq j$ and $p \in VC(p_i)$ *where $d(p, p_i)$ specifies the minimum distance between p and $p_i$ in Euclidean space.*

Figure 1.1 illustrates an example of a Voronoi diagram and its cells for nine generators. Voronoi diagrams have several geometric properties. Here we enlist their main properties that we use to establish our proposed solution.

**Property 1:** The Voronoi diagram for a given a set of generators is unique.

**Property 2:** The nearest neighbor of any query point inside a Voronoi polygon is the generator of that polygon. For example, in Figure 1.1 *p1* is the nearest neighbor of query object *q*.

**Property 3**: With VD, each generator on the average has at most 6 neighboring generators which enables local search and limit the number of candidates in proximity search.

## 1.4 Challenges

The problem of devising a scalable distributed throwaway index structure is particularly challenging for many reasons that are specific to the partitioning a dynamic dataset. First, it is essential to create equi-sized partitions; otherwise, the server storing the largest partition will become the bottleneck and slow down the index construction. Second, if the partitioning method does not preserve spatial proximity of the objects, the queries need to be forwarded to all servers to ensure accuracy, which reduces query throughput. Third, since data objects continuously move, static spatial partitioning methods, where each server is assigned to a zip-code or grid cell, will create imbalance across the servers over time. Finally, tree-based indices cannot directly be implemented in a distributed setting by simply assigning an index node to a server, because traditional top-down search unnecessarily overloads the servers near the tree root.

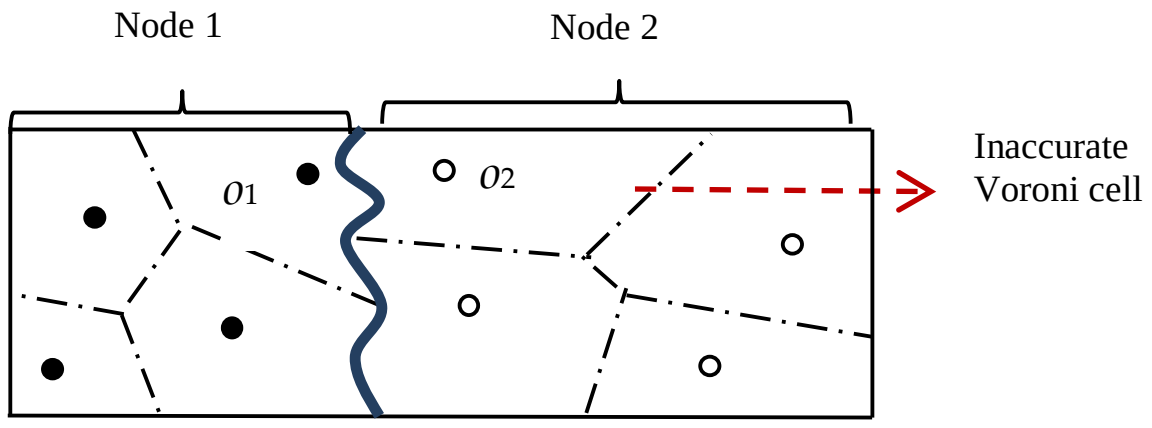

Figure 1.2: Distributed Voronoi diagram generation for 8 objects with 2 nodes

In addition to data partitioning, creating an underlying index structure that can utilize the multi-core CPUs available at server to scale up imposes great challenges as well. Voronoi diagram has a **flat** structure that lends itself nicely to parallel processing since each Voronoi cell can be built autonomously in parallel and is an extremely efficient data structure to answer a wide range of spatial queries. On the other hand, the main challenge in distributed Voronoi Diagram generation is that, due to partitioning, Voronoi cells might be inaccurate because some of their neighboring data objects may reside in a different server. For instance, Figure 3 illustrates an example of two server nodes that are asked to generate Voronoi Diagram of eight spatial objects. As shown, Voronoi cells of objects $o_1$ and $o_2$ are inaccurate. This is because even though $o_1$ and $o_2$ are Voronoi neighbors, they are sent to different nodes in which local Voronoi diagrams are created independently. One way to fix this problem is to gather all of these inaccurate cells in a single server and fix them there. Clearly, such a solution would create a bottleneck and limit the scalability of the index construction. Therefore, it is important to identify these inaccuracies at each server and fix in parallel.

In the cost-efficient partitioning problem, including the workload aspect into data partitioning phase yields new challenges. First, even though workload associated with the partitions is taken into account, the spatial proximity of the data objects still need to be preserved; otherwise, the queries will have be forwarded to all servers which reduces query throughput. Second, once a partition plan is generated for a given dataset that maximizes the server utilization, access patterns of partitions might change over time. Similarly, insertion/deletion/update of the data objects results in inclusion or exclusion of partitions. The naïve way to handle these changes is to come up with an offline partitioning approach and execute it periodically on the updated dataset and generate a fresh partition plan. The problem is that the new partition plan might require redistributing the entire dataset across the servers which results in tremendous amount of unnecessary network traffic and a very long downtime for the application.

# Chapter 2: Related Work

## 2.1 Spatial Indexing Techniques for Dynamic Datasets

### 2.1.1 Moving-Object Index Structures

There is an extensive literature on general-purpose spatial indexing techniques. Among these approaches R-tree [Gutman 1984] and Quadtree [Finkel 1974] are the most prominent index structures as they are also employed in commercial products such as Oracle. However, handling highly dynamic datasets is not the strong suit of these conventional index structures [Akdogan 2010]. Therefore, specialized moving-object techniques have been proposed [Tao 2003, Jensen 2004, Chen 2008 and Nguyen 2012] where the main idea is to update the index when an object's velocity changes. The main problems with these incremental moving object indices are as follows. First, they make the simplifying assumption that extra information about the object movement behavior is known in advance, such as direction and speed, which limits the applicability of the approach for unpredictively moving objects. Second, given the limited computational capability and storage of a single machine, a centralized system will eventually suffer from performance deterioration as the size of the dataset or the number of queries increases.

### 2.1.2 Throwaway Spatial Index Structures

In throwaway indexing model, a static index is periodically built from scratch rather than processing updates in-place. With this model, queries are forwarded to a read-only static index. Meanwhile, the latest positions of the data objects are collected in a buffer (e.g., hash table), a fresh is index created and swapped with the current read-only index. Recently, a few throwaway index structures have been proposed. TwinGrid [Sidlauskas 2011] maintains two separate grid indices, one for collecting updates and one for processing queries, where the data is occasionally copied between these indices effectively with memcpy. TwinGrid indeed processes updates; however, since the queries are forwarded to the static read-only index, which is updated periodically, we classify it as a throwaway index. MOVIES [Dittrich 2011] collects updates in a buffer (hash table) and periodically builds a linearized kd-tree using a sorted z-ordered sequence which results in *approximate* query answering. In addition, both approaches suffer from limited scalability due to their single-server design.

To address the scalability issues, a distributed version of MOVIES, which we name as D-MOVIES, is also discussed in [Dittrich 2011]. Specifically, first it hash-partitions data object by unique object ids across multiple nodes and then runs the centralized index generation algorithm at each server in parallel. With hashing, the global spatial index is not broken into disjoint sub-indices since the ranges of the local linearized kd-trees built at each node overlap with each other. Therefore, D-MOVIES adapts *intra-query* parallelism paradigm where queries are forwarded to all nodes. This form of parallelism is best suited for complex long-running queries. However, in our target applications, data objects execute queries with small ranges to sense their surroundings. Thus, query throughput degrades as the number of nodes increases due to higher query coordination cost.

### 2.1.3 Distributed Spatial Index Structures

There are few recent studies that adapt a shared-nothing distributed architecture. RT-CAN [Wang 2010] integrates a CAN-based [Ratnasamy 2001] routing protocol with an R-tree based indexing scheme to support multi-dimensional query processing in a cloud system. QT-Chord [Ding 2011] is very similar to its preceding approach RT-CAN with the main difference that it combines Chord [Stoica 2001] routing protocol with Quadtree. In response to a data update, CAN finds the target node to send information with O($\sqrt{N}$) number of network messages, where N is the number of nodes in the cluster. Considering the frequent updates from the (moving) objects and high network overhead, it is impossible to keep up with such index updates. Note that although Quadtree outperforms R-tree in terms of updates in a centralized system, in a distributed environment Quadtree is no longer more efficient than R-tree as they have almost the same network cost, which dominates the total cost; therefore, both these approaches suffer from update. RT-CAN employs *inter-query* parallelism paradigm that enable to process multiple queries in parallel resulting in higher query throughput. Several other approaches that use distributed hierarchical index structures such as R-tree based SD-Rtree [Mouza 2007] and kd-tree based k-RP [Litwin 1996] have been proposed for parallel spatial query processing. The major problem with the tree-based approaches is that they do not scale due to the traditional top-down search that unnecessarily overloads the nodes near the tree root.

Finally, Table 1 summarizes the prior work, which are most relevant to our approach, in terms of architecture, indexing model and query processing method. We compare our approach with RT-CAN and D-MOVIES in the experiments.

Table 2.1: Comparison of Relevant Approaches

| Approach | Architecture | | Indexing Model | | Querying Method | |
|---|---|---|---|---|---|---|
| | Centralized | Distributed | Incremental | Throwaway | Inter-query | Intra-query |
| RT-CAN | | ✓ | ✓ | | ✓ | |
| MOVIES | ✓ | | | ✓ | | |
| D-MOVIES | | ✓ | | ✓ | | ✓ |
| D-ToSS | | ✓ | | ✓ | ✓ | |

## 2.2 Workload-based Partitioning Techniques

While we propose a cost-efficient partitioning technique to minimize the server cost, most of the existing techniques aim to 1) co-locate frequently accessed objects together and 2) create equi-sized partitions. We discuss these partitioning techniques in three different contexts: spatial databases, relational databases and No-SQL & NewSQL databases. Then we discuss the literature on energy-efficient computing separately and conclude the section.

### 2.2.1 Spatial Databases

Most of the existing approaches are designed for a centralized paradigm where all spatial operations are performed on a single server. Among these tree-based, R-tree [Gutman 1984] and Quad-tree [Finkel 1974] are the most prominent index structures which are implemented in commercial products. Conversely, several other techniques employ a flat structure such as Voronoi diagram [Okabe et al. 2000] which decomposes the space into disjoint subsets. Likewise, the method in [Shekar 1998] converts spatial objects into a graph and divide the graph using min-cut algorithm to maximize the distance among the partitions. The main problem with these techniques is that they all suffer from workload skew as the main objective is clustering spatially close objects. In addition, tree-based methods cannot directly be applied in a distributed system as whichever server holds the higher levels of the hierarchies (e.g., root) becomes the system bottleneck [Akdogan 2014].

There are few recent studies that handle spatial data in the context of distributed systems. RT-CAN [Wang 2010] integrates a CAN-based [Ratnasamy 2001] routing protocol with an R-tree based indexing scheme to support multidimensional query processing in a cloud system. QT-Chord [Ding 2011] is very similar to its predecessor, RT-CAN, with the main difference that it combines the Chord [Stoica 2011] routing protocol with Quad-tree. Note that, both Chord and CAN adapt a storage-oriented balancing strategy which leaves approximately half of the cluster idle if the application is used in different time zones.

### 2.2.2 Relational Databases

In relational databases, there is an extensive literature in automated database partitioning. Among these, most prominent are the products from two large companies: Microsoft's SQL Server AutoAdmin [Agrawal 2003, Nehme 2011, Polychroniou 2014] and IBM's DB2 Database Advisor [Rao 2002, Zilio 2004]. These techniques take three inputs: 1) schemas of the tables, 2) stored procedures and 3) the log of previously executed queries. Subsequently, a final database design is generated based on these inputs. The main objective of these approaches is to co-locate tables, a fragment of tables or individual rows that are frequently accessed together to increase query throughput. Moreover, these methods are devised for single-server centralized systems, where distributed transactions, network I/O and server cost are not taken into consideration. Therefore, we limit this discussion to the related work in ***parallel*** *DBMSs*.

The key differences among the existing partitioning techniques in parallel DBMSs are in 1) identifying the partitioning attributes and 2) the search process used to find the optimal partitioning strategy. For example, Schism seeks to minimize distributed transactions [Curino 2010]. For a given database, Schism populates a graph containing a separate vertex for every tuple and creates an edge between two vertices if the tuples that they represent are co-accessed together in a transaction. The edge values proportionally increase with the common transactions. It then produces balanced boundaries using a graph partitioning technique that minimize the number of cross partition edges. The work in [Nehme 2011] employs a branch-and-bound algorithm to search for table partitioning and replication choices for shared-nothing, disk-based DBMSs. The main problem with these approaches is that they put co-accessed tuples into the same partition which reduces the network I/O and query coordination cost resulting in increased query throughput. In addition, these techniques only run on static datasets, where updates are not taken into account.

**Another related service is Akamai** that provides a distributed-computing platform through thousands of servers deployed around the world. Akamai geographically distributes the content and serves it to the users from the closest location with the objective of reducing the propagation delays in the network. On the other hand, our objective is different as we aim to minimize the server cost by maximizing the utilization at each server in a Cloud Computing Infrastructure. In fact, a recent study shows that if Akamai's resource consumption was completely proportional to load it handles, *40%* of the power cost could be eliminated [Qureshi 2009].

### 2.2.3 Relational Databases

Recently, several NoSQL database products such as MongoDB [Mongo 2015] and HBase [HBase 2015] provide scalable solutions by giving up on strong consistency (ACID properties). These products have *geospatial* data support as well; however, they still use similar partitioning techniques as parallel databases (e.g., hash or range partitioning). With MongoDB, dataset is partitioned into smaller chunks and scattered across the servers. MongoDB employs a static *storage-oriented* partitioning strategy, with which if too many partitions are assigned to a server, some of those partitions are migrated to other servers. The main problems with this approach are as follows. First, data migration impacts the performance. Second, it only balances the amount of the data at each server without considering if the data is accessed or not. Big data processing frameworks such as SpatialHadoop [Eldawy and Mokbel 2015] might be relevant; however, the main objectives in such systems are achieving scalability, reducing network I/O and balancing the load across servers given a long-running data processing task.

### 2.2.4 Energy-Efficient Computing

There has been a growing amount of research in green computing that aims to reduce power usage in data centers which account for about 2% of all electricity consumption in the U.S. [Chen 2008,

Joukov 2008, Lin 2013, Yao 2014]. However, there are many open problems in this field since even today's energy-efficient data centers consume a non-trivial amount of energy (almost 50% of their peak) even when they are nearly idle [Chen 2008, Lin 2013].

The most relevant literature to our problem is in right sizing of data centers where the idea is to dynamically control the number of activated servers and switch the idle servers into power saving mode [Lin 2013]. On the other hand, this power-oriented optimization problem considers numerous factors such as 1) the energy used toggling a server, 2) the delay in migrating connections, data, etc. when toggling a server, 3) ensuring SLAs (service-level agreement) are met, 4) the risk associated with server toggling, etc. Note that, these points are crucial for data centers but not for the users leasing servers.

A few energy-efficient data partitioning techniques have been proposed such as HARP [Wu 2013] which partitions the tables using specialized hardware to balance the gap between memory and CPU bandwidths resulting in power saving. Such studies are not relevant to our problem either since energy expenditure is not the concern of the users of cloud infrastructures.

# Chapter 3: D-ToSS: A Distributed Throwaway Spatial Index Structure for Dynamic Location Data

In this section, we propose a distributed throwaway spatial index that scales *near-linearly* both in index construction and query processing. The fundamental construct of our index structure, dubbed *D-ToSS* (for Distributed Throwaway Spatial Index Structure), is a Voronoi diagram (*VD*) where we adapt a Voronoi-based partitioning method and build a Voronoi cell for each data object in a distributed fashion. We selected *VD* as our partitioning technique and underlying index structure because 1) it has a **flat** structure that lends itself nicely to parallel processing since each Voronoi cell can be built autonomously in parallel and 2) it is an extremely efficient data structure to answer a wide range of spatial queries [Sharifzadeh 2010]. The main challenge in distributed Voronoi Diagram generation is that, due to partitioning, Voronoi cells might be inaccurate because some of their neighboring data objects may reside in a different server. The intuitive approach to overcome this issue is to first build the global Voronoi over all the data objects on a single server and then partition it across the servers for query processing. However, this *build-first-distribute-later* approach [Akdogan 2010] suffers from limited scalability. To overcome this challenge, we propose a novel three-step *distribute-first-build-later* scalable framework.

First, *D-ToSS* employs a Voronoi-based *adaptive* partitioning technique that quickly learns from the dataset and distributes the objects across the servers while preserving the spatial proximity among the objects and balancing load among the servers. Second, to take advantage of the multi-core architecture of each server node, we generate local *VD*s using multiple threads where every Voronoi cell is generated by a thread autonomously. Subsequently, we identify whether a generated cell at a server is globally accurate or not, without requiring any information about other servers. Then, we refine inaccurate cells effectively in parallel with multi-way message-passing (all servers communicate with each other in parallel). Note that the number of messages increases along with the number of inaccurate cells. Since our partitioning method clusters spatially close objects, only very few cells at the partition borders are inaccurate. Our partitioning approach reduces the inaccurate cells to 2.5% in comparison to the 92% inaccuracy, when we randomly distributed data objects of our real-world traffic dataset. Third, we further reduce the number of network exchanges by replicating the border points of partitions so that the points with inaccurate cells can be fixed using only the local information without requiring any server communication.

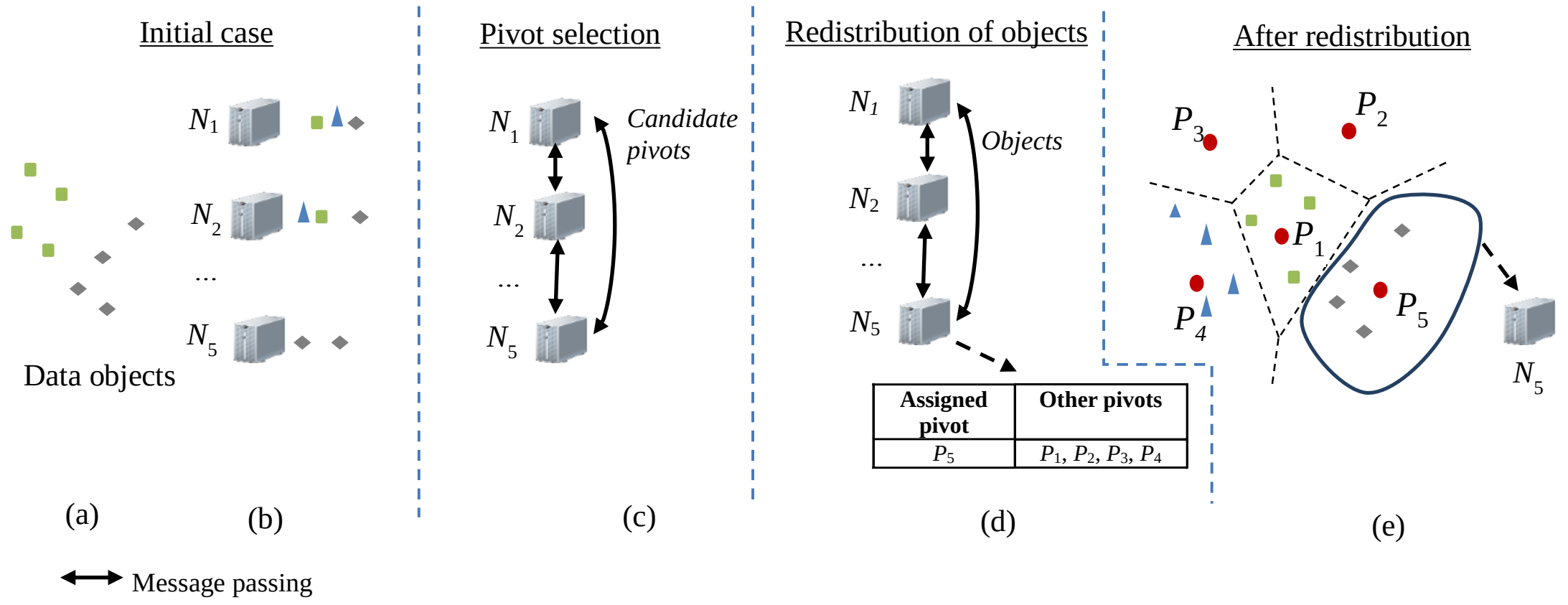

Figure 3.1 Workflow of the adaptive data partitioning phase: a) A set data objects *O* in the plane. b) Objects randomly distributed across 5 nodes. c) Pivot selection. d) Shuffling objects among the nodes based on elected pivots. Each node maintains the set of pivots. e) After redistribution objects nearby in space are stored together in the same node. Each pivot is mapped to a node.

Furthermore, to expedite query processing, we construct a hierarchical index structure on top of the local Voronoi cells at each node using multi-core CPUs. Consequently, *D-ToSS* not only can scale out because of our novel 3-step technique, but also can scale up due to our multi-core parallelization during the local index (local VD) construction. Although most of the current studies in Cloud Computing have focused on scaling out, it is crucial to utilize multi-core CPUs as chip vendors such as Intel have recently shifted to multi-core architecture rather than increasing processor speeds, which is projected to be the trend in the next 15 years [Intel 2011].

## 3.1 Parallel Voronoi Diagram Generation

### 3.1.1 Data Partitioning

Given N number of nodes and a set of data objects O which are distributed across the nodes, the purpose of this phase is to assign each object o, where $\forall o \in O$, to one of the N nodes.

The ideal partitioning should preserve spatial locality among the objects in each partition to minimize the number of inaccurate Voronoi cells (and hence message passing between the nodes) and create equally-sized partitions. Towards this end, we propose a Voronoi-based partitioning technique. The main idea is that we select *P* number of pivot data objects, where each pivot corresponds to a partition. Subsequently, we assign each object *o*, to its closest pivot. We use a two-step approach to select the pivots in a *fully decentralized* manner. In the first step, each node identifies a set of randomly selected candidate objects and broadcasts them to the other nodes. Upon receiving all candidates, every node identifies *P* number of pivots in parallel, which are *exactly the same* pivots so that nodes do not block each other. We note that each pivot corresponds to a partition and each partition (that includes subset of *O)* is stored in a separate node. D-ToSS can store more than one partition in a single node since the neighboring partitions in the same node can still preserve the spatial locality. Finally, after the set of pivots (each representing a node) are received, the nodes start to shuffle the objects by assigning each object *o* to the closest pivot (and hence its corresponding node). We formally define object & node mapping function *f* as follows:

$$f(o.location) = P_i, \text{ where } d(o.loc, P_i) = \min\{d(o.loc, P_i)\}, \ \forall \ P_i \in P$$

where o.location is composed of a latitude and longitude value, and node $N_i$ stores partition $P_i$.

Figure 3.1 illustrates an example of our data partitioning approach. As shown in Figure 3a and Figure 3.1b, the data objects are initially randomly distributed across five nodes. First, each node identifies a set of local candidate pivots and broadcasts them. Upon receiving all candidate pivots, each node computes five pivot objects (see Figure 3.1c). At this point, each node is assigned a specific pivot and also stores the information about all the other pivots. For example, node $N_5$ is mapped to pivot $P_5$ and stores the information of the other pivots $\{P_1, P_2, P_3, P_4, P_5\}$ (see Figure 3.1d). Finally, the nodes redistribute the objects based on the pivot set, where each object is assigned to the node with its closest pivot. Once the redistribution phase is completed, as shown in Figure 3.1e spatially close objects are clustered in the same node.

Next, we discuss our distributed pivot selection algorithm. As we discussed, each node $N_i$, where $\forall N_i \in N$, initially stores $O_i$ number of objects i.e., random subset of $O$, where $1 \leq i \leq N$ and $O = \bigcup_{1 \leq i \leq N} O_i$. With our approach, each node randomly selects candidate $S_i$ objects among $O_i$ and broadcasts $S_i$ to other nodes. Then, each node generates S/P (where P is the number of pivots) number of random sets where each set has $P$ objects. Then for each set, we compute the total sum of the distances between every two objects and choose the set with the maximum total sum as pivots. To generate the same sets, we adapt a deterministic randomization strategy and seed the random selection process with the same number at this phase so that every node computes the same pivots.

**Discussion:** Clearly, pivot selection and shuffling the objects accordingly help us figure out the recent data distribution which enables to 1) store spatially close objects in the same node and

2) maintain the balance across the nodes as the objects move. On the other hand, this process do not need to be repeated in every index creation cycle.

### 3.1.2 Local Voronoi Diagram Generation

In this phase, each node $N_i$ is asked to generate a local Voronoi diagram (LVD) for the corresponding objects o, where $\forall o \in O_i$. We have two specific goals in this phase: we aim to 1) exploit all multi-core CPUs available at $N_i$ to generate local VDs, and 2) effectively identify and fix the Voronoi cells which might be inaccurate because of the partitioning. We identify inaccurate cells at each server without requiring any information from other nodes and fix them in parallel. By eliminating the single merging bottleneck at this phase, we provide high scalability.

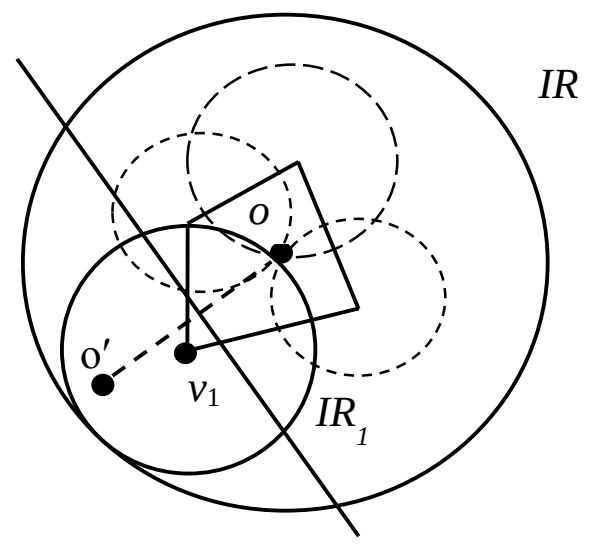

Figure 3.2: Influence region of object o.

Multi-threaded Voronoi diagram generation: A straightforward method to generate local Voronoi Diagram is to use popular sweepline algorithm [Okabe et al. 2000]. However, sweepline algorithm is *intrinsically sequential* and cannot be executed in parallel. Therefore, we break the problem into smaller pieces where Voronoi cells (VC) are generated autonomously. In particular, VC are computed by a thread in parallel and the total computation time decreases as a function of the number of cores available at each node. For instance, given a 64 core server, 64 VCs will be computed in parallel in that server. The main challenge is to create a VC independent of other VCs around it. In order to overcome this challenge, we first compute an approximate Voronoi cell for each object, which does not require any information about the other cells, and then refine it. We define these generated cells as Inaccurate Voronoi Cell (IVC). The formal definition of IVC is as follows.

*DEFINITION 2. Inaccurate Voronoi Cell (IVC)*

*Let P be a set of data points, $P^+$ and $P^-$ be subsets of P and $P^+ \cup P^- = P$. If we only consider the points in $P^+$, the generated Voronoi cells might not be accurate because there might be some points $p_j \in P^-$, which will decrease the area of the generated cells. Therefore, these cells might be inaccurate (IVC).*

Clearly, our goal is to generate an approximate cell that is as close as possible to the exact cell. We achieve this by utilizing Z-Order [Tropf 1981] space filling curve. Z-Order uses bit shuffling property to maps the elements (which are close in space) into nearby points, which preserves the spatial locality. At this point, a Z-Order value $Z(o_i)$ is computed for every object $o_i \in O$. Let $P_i^+$ be the subset of objects that will be used to construct an inaccurate cell $IVC(o_i)$ for object $o_i$. We initialize $P_i^+$ set with k nearest neighbors from the set of points $\{Z(o_{i-k}) \dots Z(o_{i+k})\}$. This is done using a linear range scan of data objects. The value of k is data-dependent; however, as every data object has at most 6 Voronoi neighbors on average (see Property 2), it is intuitive to use 3 as the k value; i.e., $P_i^+$ contains 3 preceding and following points on the Z-Order sequence; however, we determine the best k value for our dataset experimentally.

Now the challenge is how to identify the set of candidate objects that might modify an IVC. It is clear that if object o′ is far enough from another object o, it will not modify IVC(o). To identify such objects, we define an influence region (IR) for each object using the current approximate cell, which includes all possible candidate objects that might affect an IVC. Once IR is computed, we locate the Z-Order values within IR by performing range queries efficiently using bit shifting [Dittrich 2011]. Then we check the points in the result set against the current approximate cell and modify it if changes the current cell. The formal definition of IR is as follows:

Influence Region (IR) is computed based on current *IVC*. In particular, for each vertex $v$ on the *IVC* of an object $o$, the locus of the points which can exclude v from the cell is inside a circle centered at $v$ with a radius $r = d(v, o)$. Figure 3.2 illustrates the influence region of object $o$. As shown, $o'$ is inside the circle $IR_1$. The bisector crossing the line $|o, o'|$ intersects with the *IVC* causing the vertex $v_1$ being excluded from the cell. The set of circles $IR_i$ specifies the minimum influence region *IR* including all points that their presence in $P^+$ changes the *IVC*. The radius of circle centered at IR is $r_o = \max(d(o, v_i))$.

Fixing inaccurate Voronoi cells: As we discussed earlier, since the objects are divided into disjoint subsets, some of the refined VCs may be incorrect because their Voronoi neighbors reside in different nodes. To this extent, we need to identify a set of candidate objects that can possibly modify an IVC. If all these candidate objects reside in the same node, we know that the cell is accurate.

To identify inaccurate cells, we first compute, for each object o, $\forall o \in O_i$, an influence region IR(o) and then check IR(o) against the borders of the node $N_i$. If IR(o) falls inside the $N_i$, o does not need to be refined. Otherwise, o is marked as inaccurate. Note that, the inaccurate cells are the border objects of the partitions, and hence are a very small subset of $O_i$. Another important point is that we can identify such inaccuracies locally without requiring any information about other nodes and this mechanism provides high scalability because every server first detects and then fixes the inaccuracies in parallel (with multi-way message passing).

### 3.1.3 Geospatial Replication

The conventional way to replicate data is to copy a partition a fixed number of times to other nodes. However, with spatial data the points at the border of a partition are more important than the points interior to the partition. Therefore, we only replicate the border points which enables us to avoid significant amount of network communication.

Specifically, as we discussed local Voronoi diagrams in each node include inaccurate cells and we can identify them using IR. Once we identify the inaccurate Voronoi cells, we replicate their generator (data objects) across the partitions so that the potential inaccurate Voronoi cells are refined using local information. In particular, to refine IVC of o, we first identify the candidate partitions with which IR(o) overlaps. Specifically, let CN(o) be the set of nodes need to be considered to refine o and $N_i$ be the node that stores o. $N_i$ sends IR(o) to each of the nodes in CN(o) and asks for the objects within IR(o). Consequently, the IVC of o is refined based on the information passed to the node. Figure 3.3a illustrates an inaccurate Voronoi cell $o_1$ whose influence region is crossing another partition, i.e., $P_5$. To refine $o_1$, $N_1$ will communicate with $N_5$ and ask for the objects inside IR($o_1$).

Figure 3.3b illustrates our geospatial replication technique where the shaded area around partition $P_1$ is replicated into node $N_1$. In this specific case, $o_1$ does not need to be refined anymore. We note that replicating small amount of data improves the accuracy significantly. Our experiments show that by replicating 5% of the data objects, we can reduce the number inaccurate cells by 70%. The question is that how much of the border objects should be replicated to accomplish maximum accuracy. Towards that end we use range selection technique where all the objects within a certain distance $\lambda$ (see Figure 3.4 from the border of a partition are replicated. Note that, we replicate data objects at the partitioning phase by slightly modifying the object-&-partition mapping function f presented in Section 3.1.1. The formal definition of the modified mapping function is as follows:

where d(o.location, $P_i$) = min{d(o.location, $P_i$)}, $\forall$ $P_i \in$ P. Let VN($P_i$) be the Voronoi neighbors of partition $P_i$, which originally contains o. Then $P_{ne}$ = {$P_j$}, where $\forall P_j \in$ VN($P_i$), d(o, hp($P_i$, $P_j$ ) $\leq \lambda$ and hp($P_i$, $P_j$) is the hyper plane (border line) between pivots $P_i$ and $P_j$. The computation of distance is d(o, hp($P_i$, $P_j$) is presented in [Lu 2012] using triangle inequality.

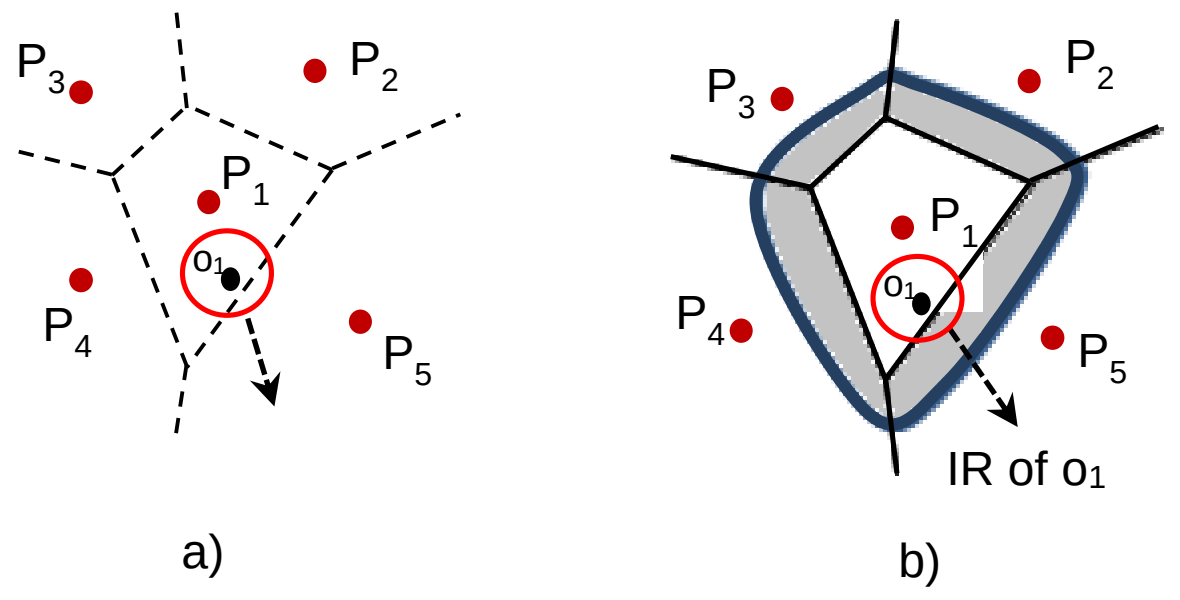

Figure 3.3: a) An inaccurate Voronoi cell (o1) whose influence region is crossing another partition (P5); therefore, o1 needs to be refined. b) A sample geospatial replication. Object o1 does no longer need refinement

Figure 3.4 shows an example of the object-&-partition mapping with geospatial replication. Partition $P_4$ is the closest pivot to object o so it is directly mapped to $P_4$. Subsequently, since d(o, hp($P_4$, $P_1$) is less than $\lambda$, it is mapped to $P_1$ as well. In this specific example, $P_{ne}$ is composed of only $P_4$; however in some cases an object might be mapped to more than two partitions. These are the objects that are at the intersection of multiple partitions.

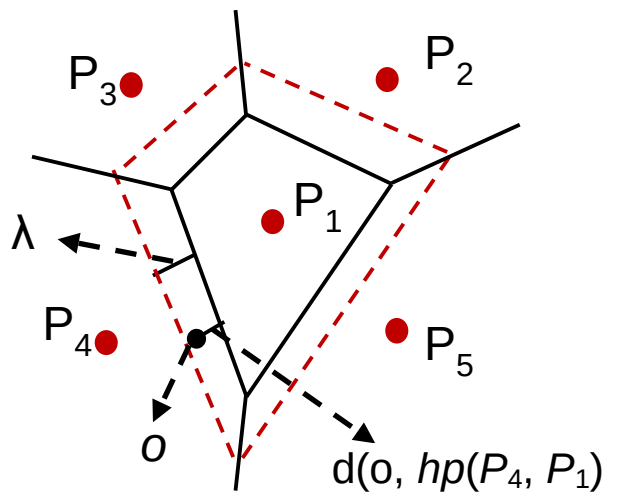

Figure3.4: All objects within $\lambda$ threshold from the border are mapped to multiple partitions

## 3.2 Building Hierarchies on Local Voronoi Diagrams

In this section, we briefly discuss how we further improve query processing by building local hierarchies on top of local Voronoi diagrams.

Several approaches have been proposed that construct hierarchies on top of Voronoi diagrams [Sharifzadeh 2010]. Rather than adapting such sophisticated techniques, we take a completely Voronoi-based approach where we organize the objects with hierarchies of Voronoi cells. A Voronoi hierarchy is defined as follows. A Voronoi cell of the higher level-i contains all Voronoi generators of the lower level-i-1 that are closer to that level-i generator than to any other [Gold 2006]. The main advantage of this approach is that we can build Voronoi hierarchies by exploiting multi-core CPUs efficiently. Our bottom-up hierarchy generation algorithm is as follow. Let *f* be the fan-out of a bucket and *c* be the number of cores available at a particular node. First, *f* number of pivots are identified, using the same technique applied in the data partitioning phase. Then each core $c_i$ is assigned *f*/*c* number of objects, and subsequently each $c_i$ maps the objects to their closest pivot. Once all objects are mapped, the same algorithm is iteratively executed for the higher level. In the following section, we show how we use these hierarchies to speed up query processing.

## 3.3 Query Processing

In this section, we show how we use *D-ToSS* to evaluate two major types of spatial queries: range and k nearest neighbor (k-NN)[1]. With *D-ToSS*, the queries can be submitted to any node in the system. This decentralization avoids unnecessary overloading of nodes unlike the tree-based approaches which overload the nodes near the root. The receiving-node either processes the query or forwards it to another node with **one network message** using the pivots we generated in the data partitioning phase.

[1] Several studies have shown the superiority of Voronoi-based index structures for various types of spatial queries (e.g., [Sharifzadeh 2010]).

In particular, let $N_{init}$ represent the server that receives query *q (that includes the query type and the location of q),* and $N_q$ represent the server that contains location of *q*. If *q* is contained in $N_{init}$ (i.e., $N_{init}=N_q$), the query is processed in $N_{init}$. Otherwise, $N_{init}$ finds $N_q$ using a random-walk algorithm on the pivot list, and then forwards *q* to $N_q$. Recall that each node maintains the list of pivots *P,* where each pivot corresponds to a node. The total cost of locating the node $N_q$ with random-walk is $O(\sqrt{P})$, where *P* is the number of pivots*,* and we consume only one network message for forwarding *q* from $N_{init}$ to $N_q$. We note that in distributed environments, the network message exchange between the nodes is the dominating cost, and *D-ToSS* minimizes this cost by using only one message. Moreover, the space requirement of storing *P* pivots at each node is very small. We store 8 bytes for the pivot location (latitude, longitude), and 32 bytes for the IP address of the host of each pivot. For example, given 1,000 pivots, the total space requirement of the pivot list is only 40KB. Once the node $N_q$ is found, *D-ToSS* processes the query using the local Voronoi Diagram (LVD) of $N_q$. In the rest of this section, we explain how we use *D-ToSS* to answer Range and k-NN queries, respectively.

### 3.3.1 Range Query

Given a dataset *O*, query point *q,* and distance *r*, range query *q* finds all the objects *O′* of *O* such that $\forall o \in O'$, $|q, o| \leq r$.

Upon receiving query *q*, $N_q$ finds the set of nodes ($IN_q$) with which *q* intersects and forwards *q* to the nodes in $IN_q$. Subsequently, each node $N_i$, where $\forall N_i \in IN_q$, executes the query in parallel. To find $IN_q$, we first include all neighbors of $N_q$ into $IN_q$ and then refine it by using *r*. In particular, let $P_i$ be one of the neighboring partitions, if $d(q, hp(P_q, P_i)) > r$, then we remove $P_i$ from $IN_q$. Otherwise, we keep $P_i$ in $IN_q$ and add $P_i$'s neighbors to $IN_q$ as potential intersecting partitions. This process terminates once all the nodes in $IN_q$ are examined.

Figure 3.5 illustrates an example of range query based on LVDs. Given a query $q_1$ with range $r_1$, we first locate pivot $P_1$ with the closest distance to $q_1$ and then initialize $IN_{q1}$ with $P_1$'s neighbors i.e., $P_2$ and $P_3$. Since $d(q_1, hp(P_1, P_2)) > r_1$, we remove $P_2$ from $IN_{q1}$. This means that there is no use for the LVD in the node corresponding to $P_2$, and hence no message needs to be sent. On the other hand, $d(q_1, hp(P_1, P_3)) < r_1$, thus we add $P_3$ and $P_3$'s neighbors to $IN_{q1}$. The algorithm runs until all the elements in $IN_{q1}$ are examined.

We can further improve the partition search algorithm by pruning false-positive partitions using the location of the furthest object (from the pivot) in that partition. In particular, even if $P_i$ satisfies the equation, $d(q, hp(P_q, P_i)) < r$, we can discard it using the following rule.

*PRUNING RULE. Furthest Object*

LEMMA-1: Let $f_i$ be the location of the furthest object in the partition from the pivot $P_i$, where $f_i = max\{|o, P_i|\}, \forall o \in O_i$. Given a query $q$ with a range $r$, even though $d(q, hp(P_q, P_i)) < r$, if $|P_q, P_i| > |P_i, f_i| + r$ then no object in $P_i$ can be in $q$'s range.

The proof is straightforward, and we explain it using Figure 3.5. As shown, $q_1$ intersects with partition $P_3$; however, by keeping the location of the furthest object $f_3$, we guarantee that $q_1$ does not cover any object of $P_3$.

Finally, once $IN_q$ is computed, we forward $q$ to each node in $IN_q$ and execute the query using the local Voronoi diagrams. For the rest of the query processing, we iteratively traverse Voronoi hierarchies using the same pivot locating and pruning strategies described above. The cost of this top-down traversal is $\sqrt{f} \cdot h$, where $h$ is the height of the hierarchy and $f$ is the fan-out. Once local query processing is completed, each node sends its partial result to the query issuer.

### 3.3.2 *k* Nearest Neighbor Query

Given a query point $q$ and a set of data objects O, $k$ Nearest Neighbor ($k$NN) query finds the $k$ closest data objects $o_{i} \in O$ to q where $d(q, o_i) \leq d(q, o)$.

The straightforward strategy to process a $k$NN query is to locate the Voronoi cell that contains $q$ and then execute the sequential $k$NN algorithm as in centralized systems [Wang 2010]. The problem with this approach is that when the value of $k$ is large, the performance degrades, especially when $q$ retrieves objects from multiple nodes. We propose an incremental algorithm, which is similar to the approach in [Maden 2007], to answer $k$NN queries using multiple nodes in parallel. The main intuition is to start the search with an estimated range, and enlarge the search range until enough number of objects is found. Thus, we express a $k$NN query as a sequence of range queries. First, we locate the node $N_q$ which contains $q$ and set an initial radius $r_{init}$ using the following equation [Tao 2004] which gives distance estimation between $q$ and its $k$ nearest neighbor in 2 dimensional space

$$r_{init} = ed(k) / k \qquad (1)$$

$$ed(k) = \frac{\sqrt{1}}{\sqrt{\pi}}\left(1 - \sqrt{1 - \sqrt{\frac{k}{O}}}\right) \qquad (2)$$

where $O$ is the total number of objects. We refer the reader to [Tao 2004] for the proofs of the above equations. Subsequently, $N_q$ locates the nodes within the $r_{init}$ range, and asks those nodes to execute a range query with $r_{init}$. Each node reports back the number of objects from their range query. If the number of objects is not enough (i.e., less than $k$), we increase $r_{init}$ by $ed(k)/k$ and continue to increase until we find k objects. Once we have enough number of objects, the nodes send these candidate objects and $N_q$ selects the closest $k$ objects to $q$ as results.

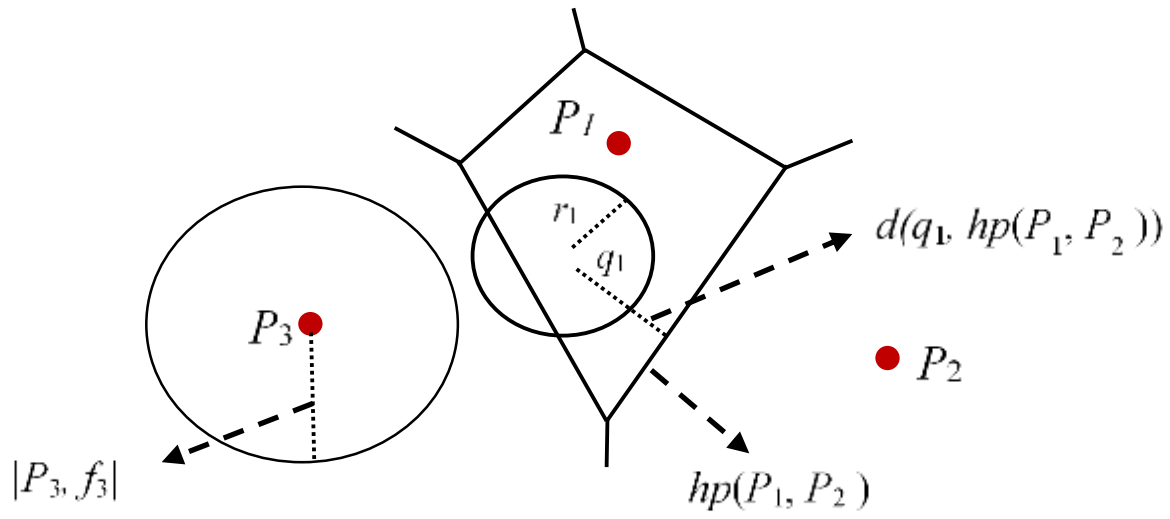

Figure 3.5: Query $q_1$ discards both partitions $P_2$ and $P_3$.

Note that even though our range and $k$NN algorithms are primitive and simple, due to the extreme parallelism, the total performance is very good. One can adapt more sophisticated algorithms and data structures that exist in the literature for processing $k$NN and range queries on Voronoi diagrams, but we believe the incremental performance benefit would not justify the complexity of implementing these algorithms.

## 3.4 Experimental Evaluation

### 3.4.1 Dataset

We use two different datasets that consists of a uniform dataset (UD) and real-world taxi dataset (TD). Our real-world TD dataset includes Beijing taxi dataset which consists of trajectories of 13,000 taxis over 3 months. Each taxi reports its location approximately in every 10 seconds on average. Since the number of objects is only 13,000 (relatively small) we split each trajectory into 25 minute intervals and treat each of these 25 minute trajectories as a different object. As a result, we generate approximately 67 million objects. For UD dataset, we also generate approximately 67 million objects (as proposed in [Wang 2010]) which are uniformly distributed in the range of $[0,10^9]$ where each object has an x and y value.

### 3.4.2 Setup and Methodology

We conduct our experiments on Amazon's EC2 cluster. We use varying number of computing nodes to evaluate inter-node parallelism (scale-out) of *D-ToSS*. The number of computing nodes in our test cluster varies from 32 to 128. We evaluated intra-node parallelism (scale-up) of our local index generation (LVD) using 8xLarge (32 CPU) nodes. These nodes are connected via 10Gbps network, and run on 64 bit Fedora 8 Linux Operating System with 60 GB memory, 32 virtual CPUs, and 4 disks with 840 GB storage. To evaluate scale-up, we disable parallel LVD generation and first run the algorithm in single-thread mode. Subsequently, we enable multi-threading mode and exploit multiple CPUs to generate LVD.

### 3.4.3 Metrics

We use the following metrics to evaluate our approach.

**Partition statistics:** The statistics about the generated partitions (e.g., average number of objects and deviation at each partition).

**Index construction time:** The total time consisting of pivot selection, data redistribution and LVD generation including Voronoi hierarchies.

**Query throughput:** Processed requests per second. We warm up the system for 60 seconds, run the experiments for 120 seconds and then compute the total number of processed queries, scale the results to 1 second and report it as throughput.

### 3.4.4 Competing Approaches

We evaluate the following approaches in the experiments.

**RT-CAN:** We first compare one-time index creation performance of *D-ToSS* with update performance of RT-CAN, where index tuning and bulk-loading of RT-CAN is enabled to enhance update performance. Subsequently, we compare the query (i.e. *k*NN and range) throughput of *D-ToSS* with that of RT-CAN, and show that our approach scales better and outperforms RT-CAN.

**D-MOVIES:** We downloaded the source code from [Dittrich 2011] where linearized kd-tree is built on Z-Order and object size is 8 bytes, and extended it where data objects are hash partitioned by object ids. Specifically, we compare the index construction time and query performances of *D-ToSS* and D-MOVIES.

Table 3.1Statistics about the Partitions for Taxi Dataset

| | | **S = 10\|P\|** | | | **S = 250\|P\|** | | |
|---|---|---|---|---|---|---|---|
| **# of partitions (P)** | **Average / Optimum** | **Min.** | **Max.** | **St. dev.** | **Min.** | **Max.** | **St. dev.** |
| 32 | 2,097,881 | 247,712 | 12,008,113 | 3,583,605 | 1,869,605 | 2,270,971 | 200,683 |
| 64 | 1,048,941 | 132,913 | 8,312,364 | 1,789,120 | 887,112 | 1,122,048 | 117,468 |
| 96 | 699,294 | 89,450 | 7,562,505 | 1,256,530 | 629,894 | 760,845 | 65,475 |
| 128 | 524,470 | 72,561 | 6,016,339 | 986,156 | 482,701 | 571,846 | 44,572 |

### 3.4.5 Effect of Partitioning

To investigate the impact of adapting a spatial partitioning method, we first randomly distribute the objects across the nodes as a baseline. For example, when we distribute the distribute the objects in TD dataset among 128 servers, 92% of the cells end up being inaccurate as spatially close objects are scattered. Even worse, we need to transfer significant amount of data over the network, about 21 times of the actual dataset, to fix these inaccuracies requiring replicating the same objects to many partitions. Moreover, in the worst case, we need to replicate the entire dataset to all nodes. Since it takes several minutes to rebuild index with this strategy, we do not provide the results.

In the next experiment, we study the impact of the number of partitions and candidate pivots on the quality of our decentralized adaptive partitioning method. Table 2 shows the statistics about the partitions generated for skewed TD dataset. *S* corresponds to the number of candidate pivots each node selects and broadcasts (e.g. $S=10|P|$ means that each node selects only 10 candidate pivots). As shown, for small *S* values, the standard deviation of partition sizes (number of objects) is high. However, as the sample size increases, standard deviation rapidly drops. Even with such small *S* values, we generate almost equally-sized partitions where the deviation is only around 10% which verifies the effectiveness of our approach. We also observe that as the number of partitions increases, deviation decreases accordingly. Therefore, another approach to have even distribution of data is generating large number of partitions and grouping them. However, in the rest of the experiments we generate one partition for each node.

### 3.4.6 D-ToSS Construction versus RT-CAN Update

In this experiment, we study the impact of varying number of nodes, CPUs, and datasets on the *D-ToSS* construction. Figure 3.6 compares the construction time of *D-ToSS* with the RT-CAN's update performance. In this experiment, we use 128 nodes and the uniform dataset randomly distributed across the nodes. Note that, we assume the objects are initially randomly distributed among the servers. Our purpose here is to test the worst case performance of *D-ToSS* construction time by increasing the number of objects needs to be transferred over the network in the shuffling phase. To update the location of the objects in the uniform dataset, we randomly generate two values x', y' within [∓100, ∓250] range and add them to the current position (x, y) in the coordinate system. The justification for [100, 250] range is as follows. According to Google maps, a human walks 80 meters/minute on average. The perimeter of the world is $40x10^6$ meters. When we convert the walking speed in real life into our coordinate system ([0, $10^9$]), a human can walk 2,000 units per minute on average. Assuming a mobile device can report current location in several seconds, a number between 100 and 250 corresponds to how much a person can walk. We simulated human behavior in our experiments with the uniform dataset. As shown, even if more than 6% of the objects issue location updates it is faster to recreate *D-ToSS* from scratch. We observe the same pattern in Figure 3.7, where we run the same experiment with the taxi dataset. The only difference is that with the taxi dataset both RT-CAN and *D-ToSS* get slower. This is because the change in the location of a taxi between two consecutive GPS readings is larger compared to the distance a person can go. With RT-CAN larger movement leads more update propagation and increases the total update cost. With *D-ToSS*, larger movement does not have any impact; however, the distribution of the real data is skewed as taxis are more clustered in some regions and skewed datasets slightly reduce the balance among the nodes in terms of the number of objects that each node stores.

To explain why a complete index construction is faster than updating 6% of objects, let us analyze each cost in details. We first analyze the update cost of RT-CAN. The performance of RT-CAN is severely affected by a 3-step index update approach. First, RT-CAN locates the node which contains the query object by redundant routing between the nodes. This routing cost is $log(N)$ messages, where $N$ is the number of nodes in the cluster. Second, it updates the local R-tree with respect to new updates. Third, since the structure of the local R-tree in the node changes, the updates are propagated to other nodes with additional $3\log(N/4)$ messages. Let us explain the index update cost of RT-CAN using our UD dataset. In particular, 6% of UD dataset with 128 nodes corresponds to 4.02 million objects (i.e. 6% x 67 million). In the first step, RT- CAN passes 7 messages (i.e., log(*128*)) to locate the node whose index needs an update. Those 7 messages are need to handle only one update, and hence the total number of messages for 4.02 million updates is $28.14x10^6$, which is already 42% of the entire dataset. Once the nodes are found, in the second phase, the local R-trees are updated, where even a single object might change the whole structure of an R-tree. In the third phase, since the local R-trees are different now, RT-CAN propagates the index changes to other nodes, which incurs additional $3\log(N/4)$ messages per each update propagation.

Now let us analyze the cost of creating *D-ToSS* from scratch. *D-ToSS* only incurs $\beta \cdot |O| + \alpha \cdot |O|$ number of network messages to re-create the index structure, where $\beta \cdot |O|$ is the total number of objects mapped to a node, and $\alpha \cdot |O|$ is the number of objects exchanged between the nodes to fix inaccurate Voronoi cells. Clearly, when the objects are randomly distributed, in the worst case every object $o$, $\forall o \in O$, is mapped to a different node; therefore $\beta=1$. Obviously, it is unlikely to have $\beta=1$ with large number of objects. As we discussed earlier, inaccurate cells are the border cells and there are very few of them. Therefore, $\alpha$ is a small decimal number.

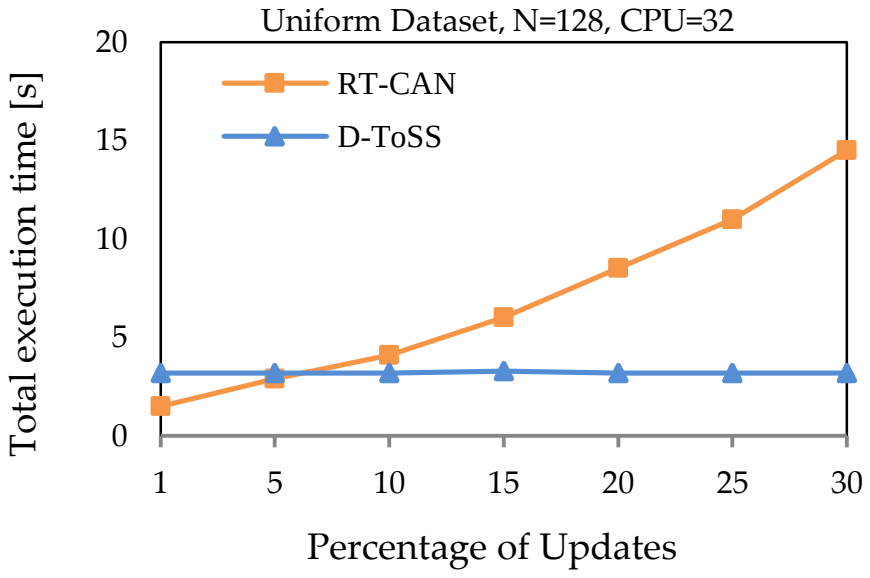

Figure 3.6: D-ToSS construction versus RT-CAN update (Uniform Dataset)

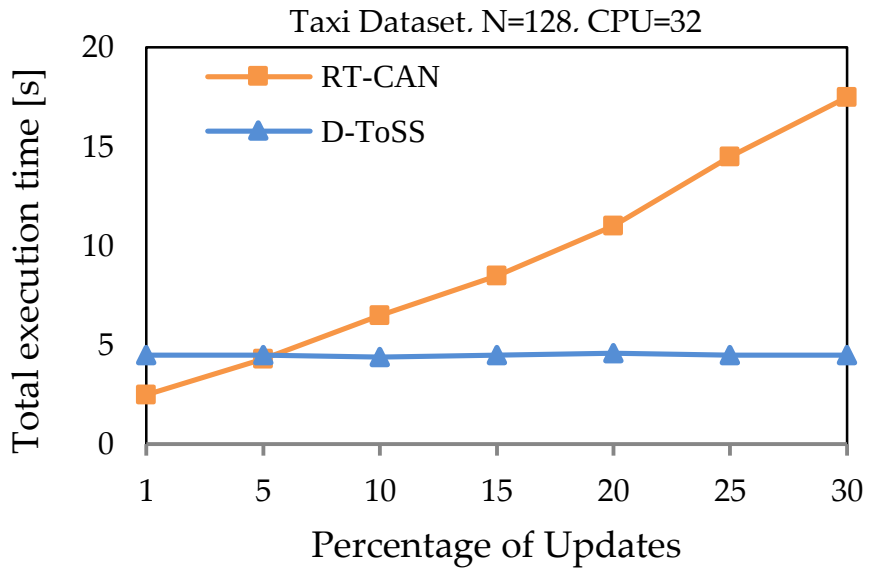

Figure 3.7: D-ToSS construction versus RT-CAN update (Taxi Dataset)

### 3.4.7 D-ToSS versus D-MOVIES constructions

Figures 3.8 and 3.9 compare the index construction of D-MOVIES and *D-ToSS* for both datasets. We vary the number of nodes from 32 to 128 where each node stores approximately 500k objects and C=1. In this specific experiment, we report the total time to generate LVDs for *D-ToSS* in the second index generation cycle, where pivot selection and data shuffling are disabled. This is because we manually hash partition the dataset by unique object ids for D-MOVIES as there is no mechanism for adapting the cluster to the current data distribution. As shown, for both datasets D-MOVIES outperforms *D-ToSS* due to its lightweight structure and hash partitioning that creates almost equi-sized partitions. However, the spatial proximity among the objects is broken with hashing. Thus, the global index is not really broken into disjoint sub-indices which requires to send the queries to all nodes which drastically reduce query throughput as shown in the next section.

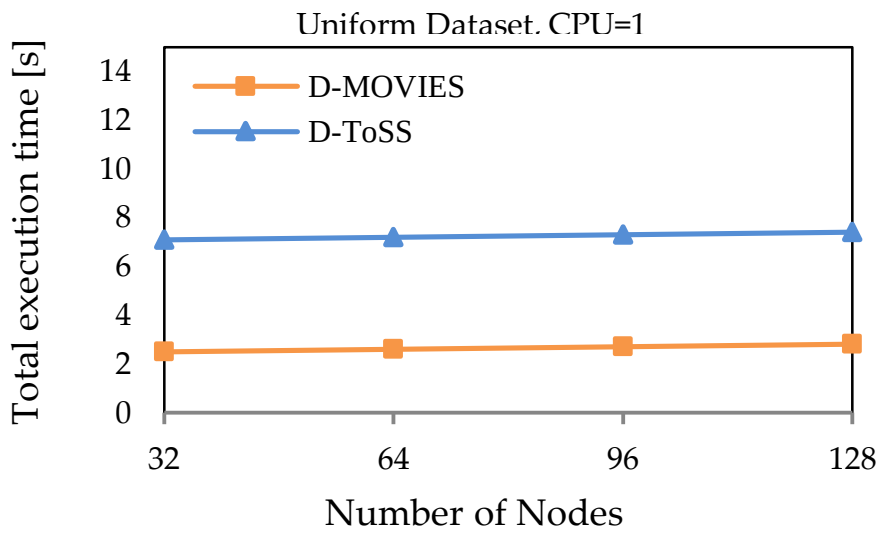

Figure 3.8: D-ToSS construction versus D-MOVIES. (Uniform Dataset)

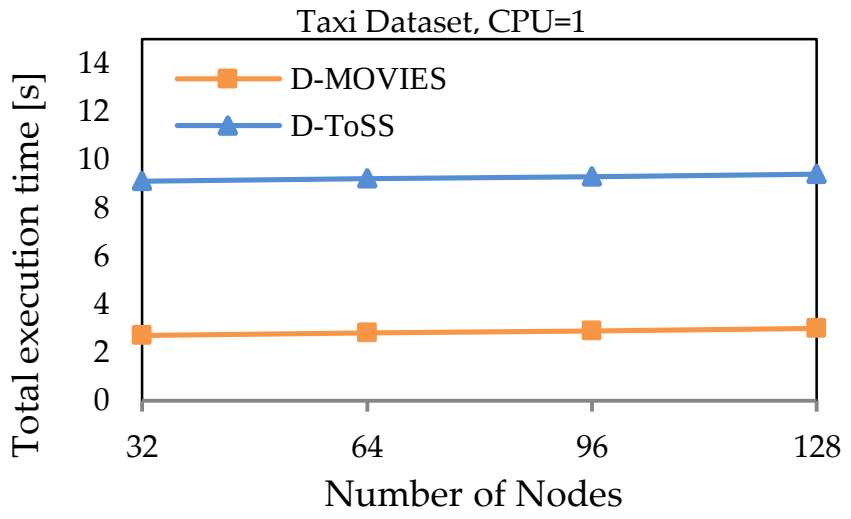

Figure 3.9: D-ToSS construction versus D-MOVIES. (Taxi Dataset)

### 3.4.8 Scale-out

Figure 3.10 shows the scale-out performance of D-ToSS. In this set of experiments, we vary the number of nodes in the cluster from 32 to 128, and evaluate the amount of time required for index creation using both uniform and taxi datasets. In distributed systems, scale out performance is evaluated by fixing the amount of data at each node and increasing the number of nodes [Wang 2010]. Note that the total amount of data still increases as more nodes are employed. In this specific experiment, each node contains 500k objects, where objects are randomly selected for TD dataset. As shown, the total execution time is almost constant which verifies the scalability of D-ToSS construction. We observe that there is an insignificant increase in the total execution time as we employ more nodes. This is because the total number of candidate pivots increases as more nodes are used (recall that nodes select a fixed number of candidate pivots) and hence it takes a little longer for the master node to identify pivots.

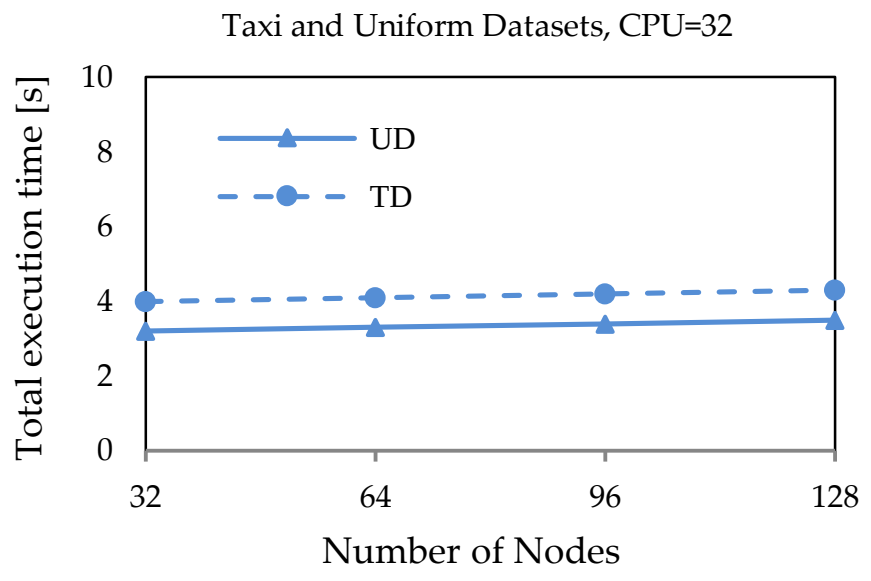

Figure 3.10: Scale-out performance of D-ToSS construction

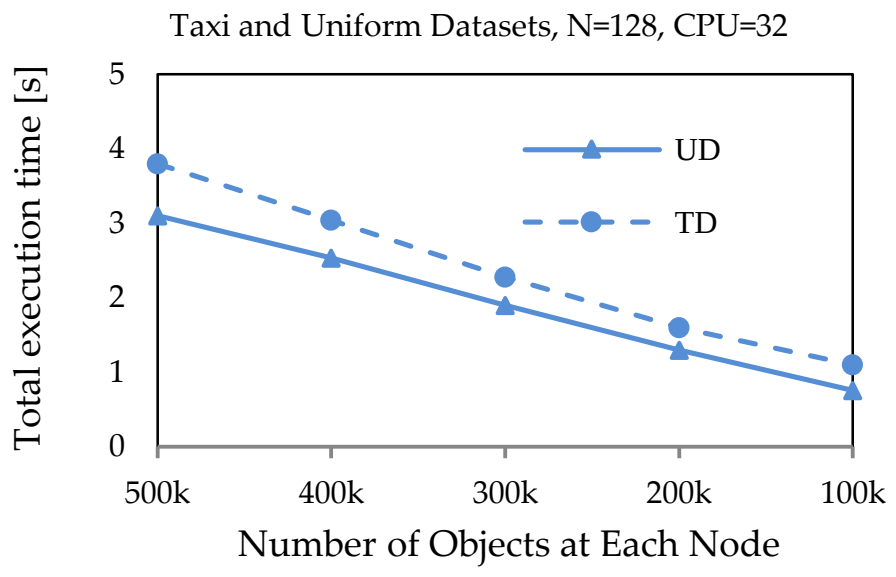

Figure 3.171: Effect of Object number on D-ToSS construction

### 3.4.9 Effect of Number of Objects at Each Node

Figure 3.11 illustrates the impact of data size on the index construction time. We fix the node number (N=128) in this experiment, and vary the number of objects at each node from 500,000 to 100,000 where objects are randomly selected. As shown, the total time almost linearly decreasing as we store less objects at each node. For the applications requiring low-staleness such as flight control systems working, *D-ToSS* can be configured accordingly.

### 3.4.10 Scale-up

Figures 3.12 and 3.13 illustrate the effect of varying the number of CPUs at each node when constructing *D-ToSS* for uniform and taxi datasets, respectively. Again we assume objects are randomly distributed at the beginning to test the worst-case performance of shuffling phase. With *D-ToSS*, since the number of network messages are minimized, the total index generation time is dominated by the LVD generation phase. Therefore, we enable multi-threading mode and generate LVDs using intra-node parallelism. As shown in both figures, as the number of CPUs increases, the LVD construction time improves significantly. We observe that the uniform dataset scales-up better than the taxi dataset. The main reason behind this difference is as follows. The LVD generation consists of two steps: Voronoi cell generation and refining inaccurate cells. With the uniform dataset, geospatial replication eliminates almost all of the inaccurate cells. Therefore, once the Voronoi cells are generated, the algorithm terminates. As there is no inter-node communication to refine Voronoi cells, the impact of increasing the number of CPUs is more significant. On the other hand, with the taxi dataset, by exploiting multiple CPUs we generate Voronoi cells faster; however, the inaccurate cells need to be refined in the second step. The inter-node communication becomes the dominant cost as we reduce the construction time of the Voronoi cells by exploiting multiple CPUs.

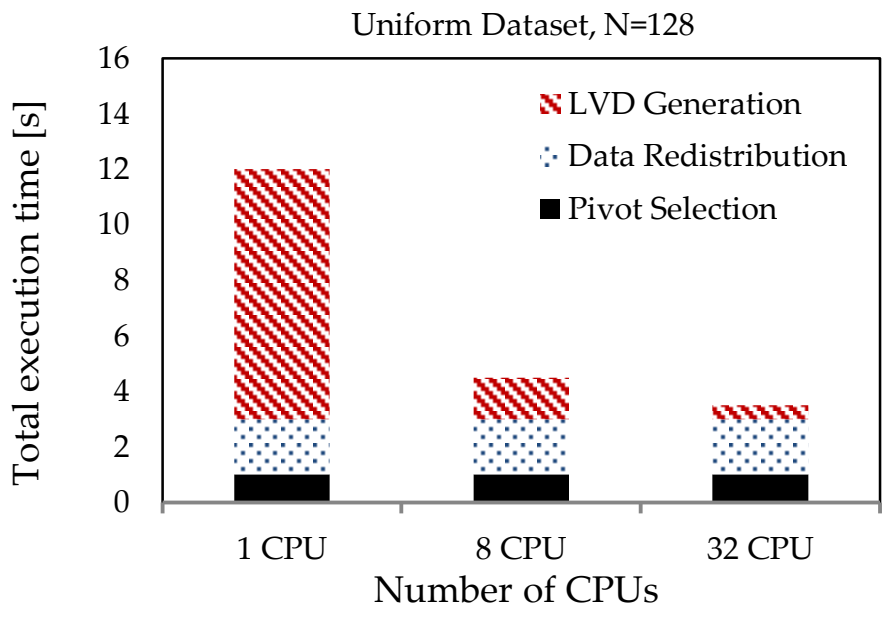

Figure 3.12: Scale-up performance of D-ToSS construction (Uniform Dataset)

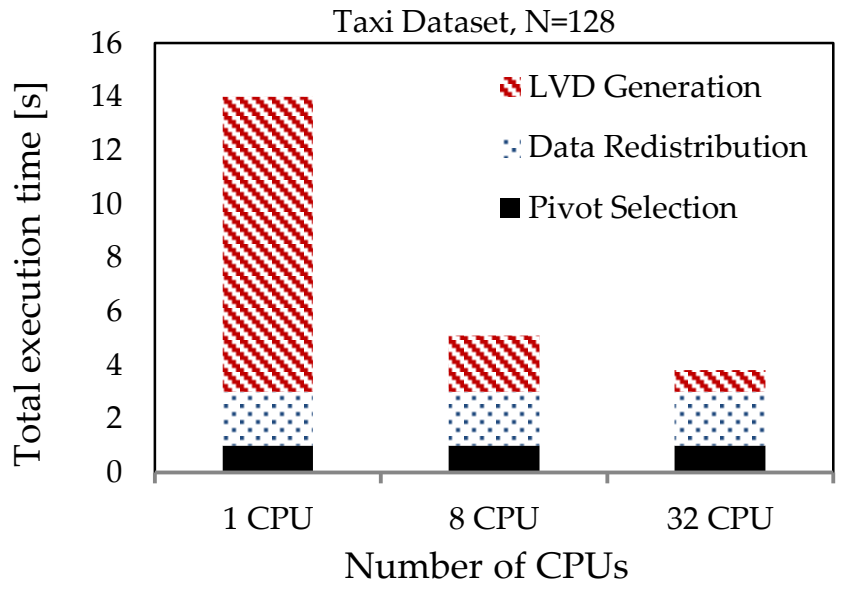

Figure 3.13: Scale-up performance of D-ToSS construction (Taxi Dataset)

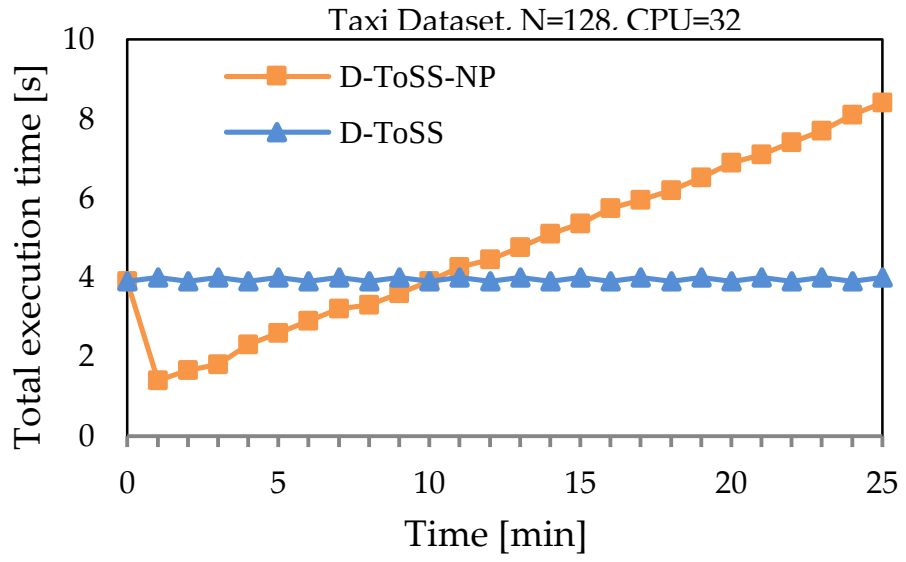

Figure 3.14: Effect of Re-partitioning on D-ToSS construction

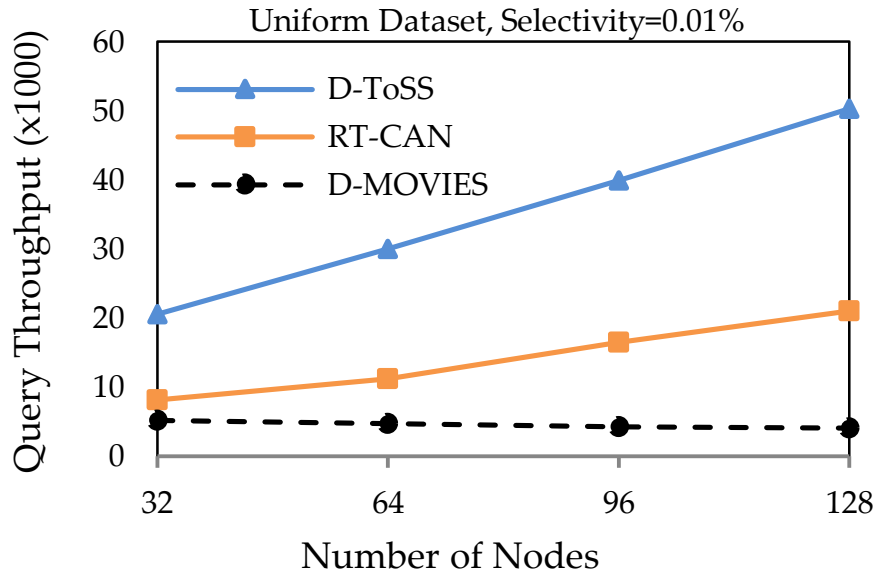

Figure 3.15: Effect of Number of Nodes on Range Query (Uniform Dataset)

### 3.4.11 Effect of Re-partitioning

Re-partitioning, which consists of pivot selection and object redistribution, is not necessary at every index construction cycle especially if the objects move slowly. Figure 3.14 illustrates the impact of performing repartitioning at every index construction cycle on TD dataset where taxis move on their trajectories for 25 minutes, N=128 and C=32. NUTS-NP distributes the objects to the nodes only once and skips repartitioning in the remaining cycles. In this experiment, we let the objects move between the partitions and the shape of the partitions do not change. This is the setting for static partitioning. As shown, NUTS-NP is slightly better at the beginning; however, after about 10 minutes, it becomes more efficient to repartition the dataset. This is because as time goes on, taxis move to other partitions and nodes suffer from unbalanced workload. We also observe that repartitioning ensures that spatial proximity among the objects is maintained at each node; therefore, it takes shorter time to generate LVDs. Consequently, when the objects move slowly, data partitioning phase can be executed less frequently. Such a strategy decreases the total execution time significantly and the total time to rebuild D-ToSS will be approximately 2% (See Figures 3.9 and 3.13).

### 3.4.12 Range Query

In this set of experiments, we evaluate the performance of range queries. We define range (with respect to a query point) as the percentage of the entire space, and as proposed in [Wang 2010] we consider three separate selectivity ranges: *small* (0.01%), *medium* (0.05%) and *large* (0.10%). We randomly generate the query points and send each query to a *D-ToSS* node in a round-robin fashion.

Figure 3.15 compares the throughput performances of *D-ToSS*, RT-CAN and D-MOVIES running small selectivity queries on varying number of nodes. As shown, *D-ToSS* outperforms other approaches in all cases. We observe that, with D-MOVIES, query throughput decreases as the number of nodes increases. This is because the queries forwarded to all nodes, and thus the query coordination cost increases due to high number of servers involved in query processing resulting in performance degrades. Another important observation is that as we employ more nodes in the cluster, *D-ToSS* shows a better linear increase in the query throughput. The reason is that with RT-CAN, increasing the number of nodes results in a longer lookup path due to log(N) routing messages. QT-Chord has also a logarithmic routing cost similar to RT-CAN. As both of these approaches have similar query performances and *D-ToSS* clearly outperforms them due to its O(1) routing cost, in the rest of this section, we only present the query performance results of *D-ToSS*.

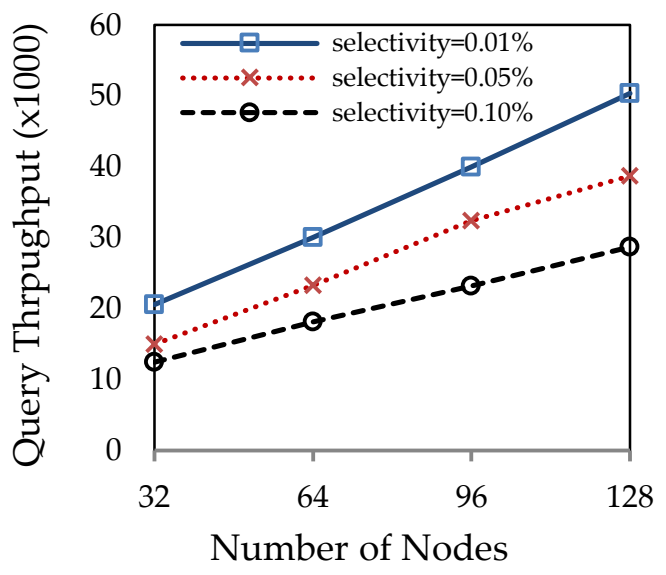

Figure 3.16: Effect of Selectivity (Uniform Dataset)

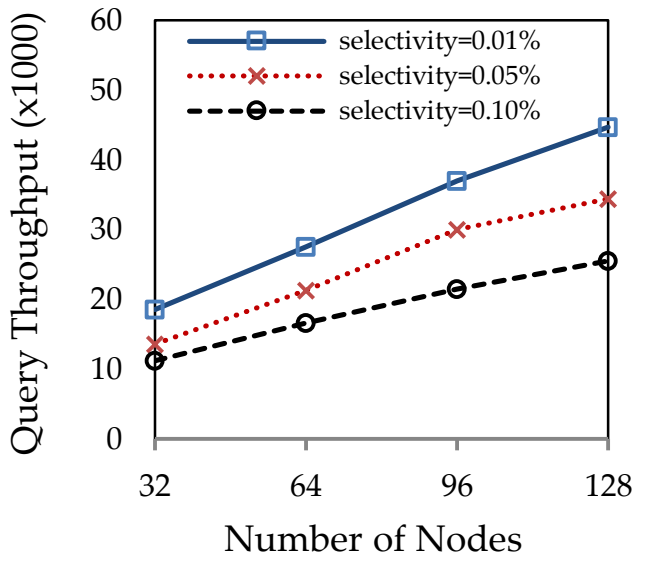

Figure 3.17: Effect of Selectivity (Taxi Dataset)

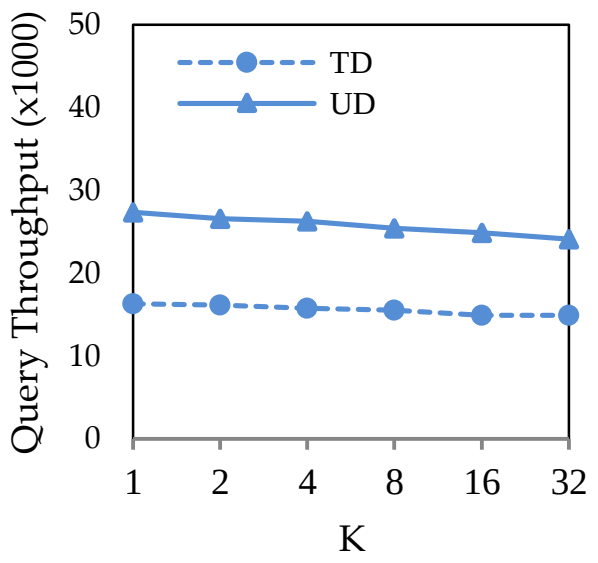

Figure 3.18: Effect of k (Both Datasets, N=64)

Figures 3.16 and 3.17 illustrate the effect of data distribution on range query for varying selectivity. We observe that as we use more nodes, the throughput increases linearly, which verifies the scalability of *D-ToSS*. The reason behind linear scalability is that the number of nodes has no impact on the number of network messages we send to process a query. With our approach adding more nodes only increases the cost of locating the node $N_q$ (the node that contains *q*) and the partition search step executed at the beginning of query processing at each node, which are insignificant. As expected, as the selectivity increases, the query throughput decreases. This is because even though larger ranges are processed by multiple nodes in parallel, we need to retrieve more objects which degrades the performance. We also observe that increasing the selectivity has marginally more impact on the query throughput with the taxi dataset. This is because with the taxi dataset, there are more partitions in the areas where the data is clustered. The queries searching for the clustered regions need to examine more partitions which results in extra message exchanges.

### 3.4.13 kNN Query

In this set of experiments, we evaluate the performance of *k*NN queries. We randomly generate the query points and send each query to a *D-ToSS* node in a round-robin fashion. Figure 3.18 illustrates the impact of the value of k on the query throughput. In this specific experiment, we used 64 nodes, and varied the value of k from 1 to 32 and compare the query throughput of *D-ToSS* with skewed and uniform datasets. As shown, the performance is less for the taxi dataset; on the other hand, the effect of k is also less compared to the uniform dataset.

## 3.5 Summary of Chapter

In this chapter, we introduced a novel distributed throwaway index, *D-ToSS*, that partitions and indexes a dynamic dataset using Voronoi diagram which we also use for query processing. *D-ToSS* not only scales out to multiple servers but also scales up since it fully exploits the multi-core CPUs available on each server. *D-ToSS* yields a very high query throughput, scales near-linearly with the number of nodes in the cluster and the total execution time remains almost **constant** as the data size increases. For example, we show 25x and 4.5x speedups over D-MOVIES and RT-CAN in query processing on a small cluster, respectively. As the cluster size increases, the performance gap between D-ToSS and others grows significantly. Moreover, our experiments show that if 2+% of the objects issue update to the index, it is faster to rebuild *D-ToSS* than update RT-CAN, when the objects are well clustered. In addition, with *D-ToSS*, a query can be forwarded to any node. This decentralization avoids unnecessary overloading of nodes that occur with the tree-based approaches. There are a few extensions for this work. First, we examine the workloads associated with each partition to maximize the server utilization which yields less number of servers, thus saving cost. All these extensions will be addressed in the next chapter.

# Chapter 4: Workload-based Partitioning of Spatial Data on Cloud

In addition to *scalability* and *efficiency*, we also aimed to maximize the server utilization that can support the same workload with less number of servers. Server utilization is a crucial point while using cloud computing because users are charged based on the total amount of time they reserve each server, with no consideration of utilization. Therefore, it is essential for users to fully utilize all servers to reduce the cost. Clearly, one key factor that impacts server utilization is the partitioning method especially in data-driven location-based services. This is because if the data partitions are not accessed, the servers storing them remain idle but the user is still charged. Whereas, existing spatial data partitioning techniques aim to 1) cluster spatially close data objects in the same node to minimize disk I/O and 2) create equi-sized partitions. On the contrary, the objective is different for cloud given the current pricing models. In this chapter, we present a novel cost-efficient partitioning method for spatial data where an increase in the servers' utilizations yields less number of servers to support the same workload, thus saving cost. Our approach considers both 1) spatial proximity among the data objects and 2) access patterns (workload) associated with the objects (e.g., how many times an object is accessed at a particular time). Extensive experiments on Amazon EC2 infrastructure demonstrate that our approach is efficient and reduces the cost by up to 40%.

## 4.1 CEPS: Cost-efficient Partitioning

In In this section, we present our workload-based partitioning approach. Given a dataset $D$ consisting of spatial data objects in 2D space and $S$ number of servers with capacity $\theta_k$ ($k \in N$),

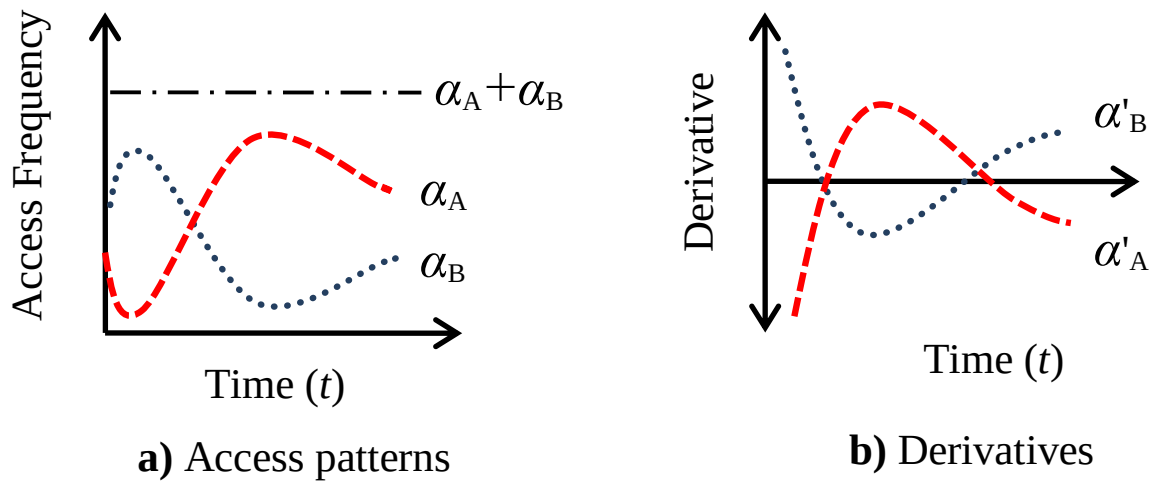

Figure 4.1: a) Two diverse access patterns A and B with workload skew. b) Derivatives of A and B (A', B'), which have a 0 sum, ensuring maximum diversity

we would like to divide *D* into *N* disjoint partitions where each partition is assigned to a server, has a flat (uniform) access pattern and fully utilizes its server. This optimization problem manifests itself into an NP-hard problem [Karmarkar and Karp 1982], with a solution space of $N^O$ where *O* is the number of objects in *D*. In addition, spatially close objects need to be stored together which adds a new dimension to the problem and makes it even more complicated. To overcome these challenges, we propose a *three* step approach as follows.

In the **first** phase, we split the dataset into *p* number of small partitions based on spatial proximity using PR kd-tree [Orenstein 1982] where $p < S$ (*p* is much smaller than *N*) and each *p* consists of certain number of objects (e.g., 10,000). We would like to note that we can also use Voronoi Diagrams as the partitioning method at this phase; however, for clarity we will use PR kd-tree in the rest of this section as it has a simpler structure. Generating such small partitions is sufficient to group co-accessed data objects together that reduces the local disk I/O at each server during query processing. For example, in real-world applications such as Yelp, users only query a limited region around them (e.g., show me the restaurants within 5 miles).

In the **second** phase, we represent each partition *p* with its access pattern $\alpha_p(t)$ at time *t*. Subsequently, we generate *N* number of super partitions (cluster) using an agglomerative hierarchical clustering (*AHR*) approach that combines partitions into clusters, where each cluster $c_k$ corresponds to a server $k \in N$. Now the challenge is how to mitigate the spikes and create uniform clusters with uniform workloads at each cluster. To this extent, we group complementing (diverse) patterns in the same cluster. As shown in Figure 1(a), diverse patterns $\alpha_A$ and $\alpha_B$ form a flat pattern when summed up ($\alpha_A$+$\alpha_B$) and the sum of derivatives (see Figure 1b), $\alpha'_A(t)+\alpha_B'(t)$, a measure of their *diversity*, is *0*. Therefore, we define a *flatness* metric $\delta$, which is the cumulative slope (derivatives) of the patterns, to identify diverse patterns and maximize diversity within each cluster. The local objective specific to the cluster $c_k$ is to minimize the *flatness* metric $\delta_k$, which is defined as:

$$\delta_k = \left| \sum_{i \in c_k} \frac{d\alpha_i(t)}{dt} \right| \quad (1)$$

In the **third** phase, we improve the solutions generated in the previous phase. The problem with *AHR* is that it might get stuck at *local optima*. More specifically, minimizing $\delta_k$ for each cluster $c_k$ itself only achieves a balanced workload on the corresponding server. However, we also aim to maximize utilization or in other words, minimize net idle-time. The idle-time $\tau_k$ for the server *k* is defined as the server time unutilized and is given by:

$$\tau_k = \theta_k - \sum_{i \in c_k} \int_t \alpha_i(t) \quad (2)$$

Although *AHR* aims at maximizing diversity of the access patterns within a cluster and does not focus on improving the global fitness metric, the constraint of server capacity drives towards a balanced utilization across servers. In order to maximize utilization, idle-time across all servers is minimized. The global fitness metric for the optimization problem, *T,* is computed as:

$$T = \sum_{k \in S} \tau_k \quad (3)$$

*Tabu search* prevents the case of a local minimum being considered an optimal solution by ensuring that a series of ensuing successive iterations do not improve the solution. In addition, it reduces the search space by avoiding swapping those access patterns which did not yield improvements to the fitness metric during neighborhood computation. In the following sections we will discuss our clustering and tabu search techniques, respectively.

### 4.1.1 Agglomerative Hierarchical Clustering

The algorithm (Algorithm 4.1) starts with an input set of $p$ singleton clusters, where each cluster $c_i$, where $0 \leq i < p$, corresponds to its access pattern. The clusters are iteratively merged to form $N$ resulting clusters. During each merge operation, two clusters $c_i$, $c_j$ are merged if the flatness metric $\delta_m$ for the cluster $c_m$ formed by merging $c_i$, $c_j$ is the minimum among all feasible pairs, i.e. $\delta_m = min(\delta_{pq})$, where $p$ and $q$ are clusters. A pair $c_i$, $c_j$ is deemed feasible if the combined access frequency of $c_i$, $c_j$ at any time is lesser than the capacity of a server. This hierarchical merging of clusters is iteratively performed until $N$ clusters are obtained. This greedy approach produces clusters that satisfy local constraints by merging diverse access patterns. However, the solution could be further improved to obtain a reduced net idle-time by improving the global fitness metric.

### 4.1.2 Tabu Search

The clustering $C_S$ of $N$ clusters produced by Algorithm 4.1 is used to initiate the Tabu search algorithm (Algorithm 4.2). This algorithm improves the current best solution by exploring its neighborhood in the solution space for better solutions subject to certain tabu criteria. Improvement made to a solution is determined by the global fitness metric $T$ (see Equation 3). The neighborhood of a solution is computed by applying local point-wise transformations to the current solution to generate a set of candidate solutions that constitute the neighborhood. Those

**Algorithm 4.1:** Agglomerative hierarchical clustering (*AHR*)

**Input**: Set of $p$ singleton clusters $C = \{c_i : \alpha_i(t) \in c_i$ and $i \in p$, set of all datasets$\}$, desired number of clusters $S$
**Output**: Set $C_N$ of $N$ resultant clusters

*num_clusters = N*
$C_N = C$
*pf = compute_pairwise_flatness*($C_N$)
**while** *num_clusters* > *k* **do**
    $(c_i, c_j)$ = *get_minimum(pf)*
    **if** *exceeds_capacity*($c_i$, $c_j$) **then**
        $(i, j)$ = *get_minimum*($P$ – *max*($c_i$, $c_j$))
    **end if**
    $c_m$ = *merge_clusters*($c_i$, $c_j$)
    $C_N = C_S + c_m - c_i - c_j$
    *pf = compute_pairwise_flatness*($C_N$)
    *num_clusters = num_clusters* – 1
**end while**
**return** $C_N$

transformations that have been applied in the recent past are added to a memory structure referred to as the tabu list, and are avoided in computing newer solutions. Maintaining the tabu list prevents cycling through solutions visited in the recent past and being limited to local optima. The current solution is replaced by an improved solution when the global fitness metric improves subject to the criteria of tabu search. Below we first describe the important aspects of the algorithm namely neighborhood, tabu list, aspiration and termination conditions. Then, we explain the entire flow of our approach in Figure 4.2.

**Neighborhood:** The neighborhood of a solution $s$ is defined as the set of solutions $S_l$ obtained by performing move transformations on a subset $l$ of fragments. A move transformation on a solution $s$ corresponding to a fragment $f$ is defined as moving the fragment from its assigned server in solution $s$ to another randomly chosen server, resulting in a solution $s' \in S_l$, the neighborhood of $s$. The set $S_l$ is selected such that the flatness metric of the clusters that are a part of this operation, changes by less than $\eta$% where $\eta$>0.

**Tabu List:** A solution $i \in S_l$, the neighborhood of the solution s, is selected to be the current solution if its fitness metric $T_i$ is an improvement from that of the current solution $T_s$, i.e., $T_i < T_s$

**Algorithm 4.2:** Tabu search

```
Input: Set C_N of N clusters
Output: Set C_N of N resultant clusters of improved
fitness metric T
tabulist = null
best_solution = C_S
current_solution = C_S
while not stopping_criterion
    neighbours = neighborhood(best_solution)
    for neighbor in neighbors
        if move(neighbor) in tabulist then
            if T (neighbor) ≤ T(best) then
                continue
            else
                best_solution = neighbour
        if T (neighbour) > T(best) then
            current_solution = neighbor
        else
            add_to_tabulist(move(neighbor))
        if sizeof(tabulist) > max_size then
            pop_tabulist()
    end for
end while
return best_solution
```

for this minimization problem and the move transformation that results in *i* is not part of the tabu list. A move *m* is added to the tabu list if the solution that is a consequence of *m* is selected to be the current solution. A move is dropped from the tabu list when the size of the list is greater than a pre-defined value on a first-in-first-out basis.

**Aspiration:** The tabu list is employed to prevent the exploration of the solution space being limited by local optima. However, this may prove to be an undesired action when the global optimal solution is already found. To prevent ignoring the global optimum, an aspiration condition is added to override the tabu list when the new solution found is better than the best solution discovered.

**Termination:** The stopping criterion for the algorithm is the failure to produce a feasible improved solution in a series of $s$, $s \in \mathbb{N}^+$, consecutive iterations after an improvement.

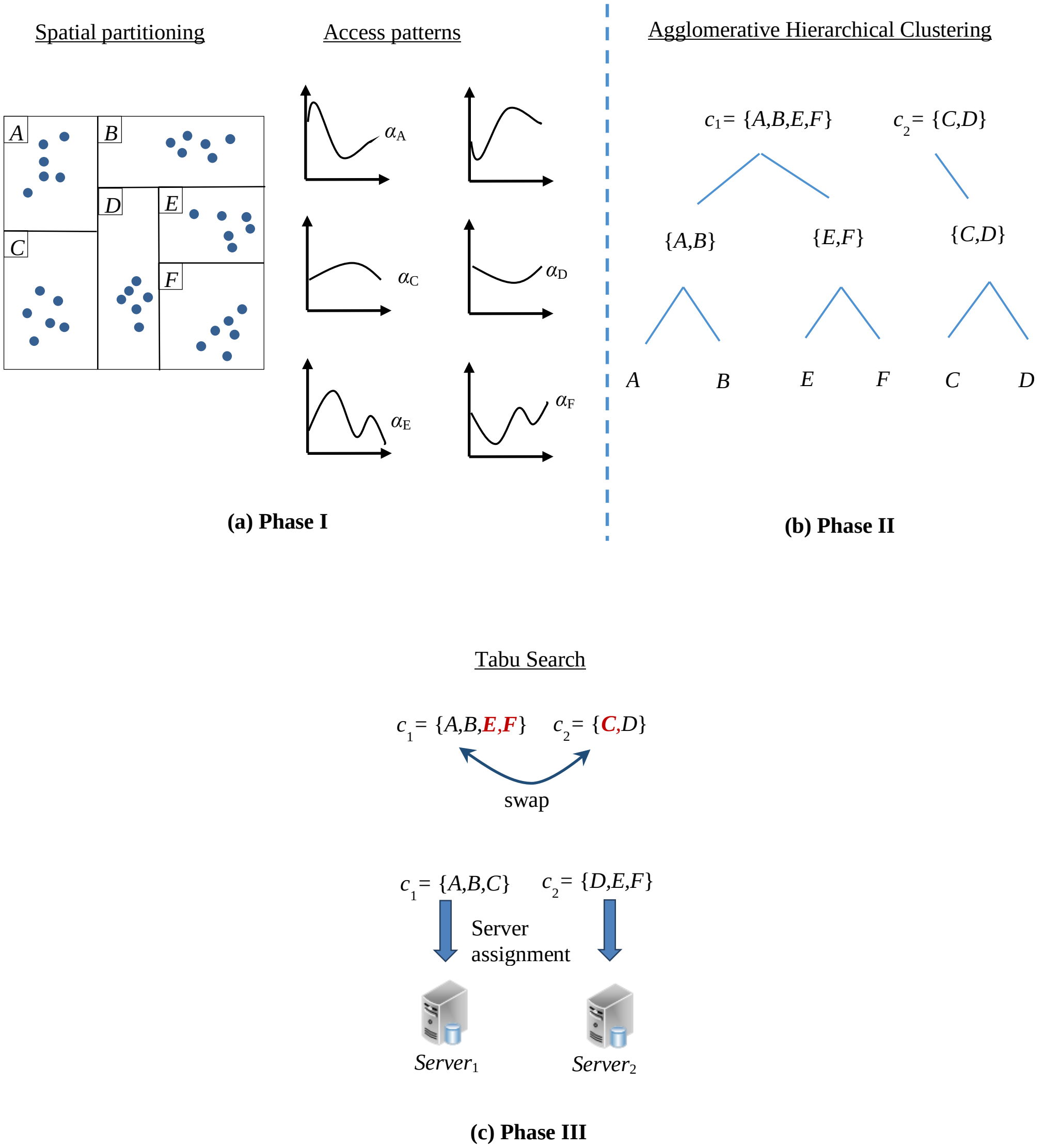

Figure 4.2: Workflow of diversity maximization, where steps a, b and c are executed sequentially. a) Spatial Partitioning of data objects which generates 6 partitions (A-F) and one access pattern for each partition. b) Bottom-up clustering phase of the three-phase approach to form 2 clusters. c) Tabu search to improve the clustering obtained in phase II, and the final partitioning.

Figure 4.2 illustrates the workflow of our approach where the data objects in 2D space are distributed across two servers, $N$=2. First, the space is divided into six partitions {$A$, $B$, $C$, $D$, $E$, $F$} using PR kd-tree [Orenstein 1982]. Then, an access pattern is produced for each partition {$\alpha_A$, $\alpha_B$, $\alpha_C$, $\alpha_D$, $\alpha_E$, $\alpha_F$}. Subsequently, *AHR* adapts bottom-up approach to hierarchically merge clusters by pairing the most diverse patterns. It returns two clusters as $c_1$={*A*, *B*, *E*, *F*} and $c_2$={*C*, *D*}. Although this assignment minimizes the workload imbalance in utilization, it does not fare well with respect to net utilization of the servers. Algorithm 4.2 searches the neighborhood of this clustering to find an improved solution. It finally returns a solution with two clusters as $c_1$={*A*, *B*, *C*} and $c_2$={*D*, *E*, *F*}, which yields the best net-utilization of servers.

## 4.2 Incremental CEPS

In this section, we discuss an incremental version of our cost partitioning approach ***(ICEPS)*** that minimizes the modifications made to the exiting clustering to reduce network I/O and downtime of the application.

*CEPS* computes a partitioning plan (clustering) for a given dataset; however, initial partition plan might need to be modified in two situations. First, insertion, deletion and update of data objects might change existing access patterns, or even inclusion and exclusion of partitions. Second, existing access patterns might change over time due to changes in user behavior over time. The naïve way to handle these changes is executing *CEPS* periodically. However, the problem is that the new partition plan might require redistributing the entire dataset across the servers which results in tremendous amount of unnecessary network traffic and a very long downtime for the application. To address this issue, in the rest of this section, we briefly discuss how we handle modifications on the dataset, and subsequently present our incremental approach to update the partition plan that minimizes the modifications made to the exiting clustering.

### 4.2.1 Insert, Delete and Update Data Objects

A data object can send an *INSERT*, *UPDATE* and *DELETE* request to the cluster, where the request includes the location of the object. Let $S_{init}$ be the server that receives the request. Upon receiving the request, $S_{\text{init}}$ forwards it to the corresponding server $S_{\text{req}}$ whose spatial zone covers the location of the object using routing tables available at each server (see Section 5 for more details about routing). Then $S_{\text{req}}$ processes the request locally using the PR kd-tree [Orenstain 1982] algorithms. Clearly, some requests will trigger merge and split operations on the partitions. Along with the partitions, new access patterns are generated or the existing ones are either removed or updated. In the following section, we discuss how the partition plan is maintained as the dataset changes.

### 4.2.2 Update Partition Plan

In this section, we discuss how we add, delete and update partitions, respectively. Algorithm 3 describes the addition of an incoming partition to the existing clustering. An incoming partition is assigned to a suitable cluster by the algorithm, which performs the operations of *grouping* and *assignment* to achieve this. Clusters are grouped on the basis of available capacity in the *grouping* operation (lines 5-8). The incoming partition is assigned to a suitable cluster among the clusters in the matched group as determined by the *assignment* operation (lines 10-17). These operations are described below.

*Grouping*: The clusters are classified into groups on the basis of the available capacity on a server represented by the cluster. The number of groups is provided as an input parameter. As a result of this operation, clusters with similar available capacities are grouped together.

*Assignment*: The capacity on a server requested by an incoming partition is used to determine the group that the partition corresponds to. The partition is matched to the group of clusters that belong to the same group as the incoming partition. Essentially, the partition is matched to the group of clusters that have an available capacity similar to the capacity requested by it. The partition is assigned to a feasible cluster $c$ that produces the minimum flatness metric $\delta_c$ among the matching group of clusters. If no such feasible cluster is found, the partition is matched to the next larger group. This process is repeated until a feasible cluster is found.

Given an incoming partition *p*, which must be added to the clustering $C_k$ of *k* partitions, the algorithm takes the number of groups, *numGroups*, required in the grouping operation as input. Grouping operation groups the clusters into *g* groups based on the capacities available on the servers represented by them and populates *G*, the map between group numbers and groups of clusters. A group number *g* is assigned to the partition *p* based on the server capacity requested by it. *p* is allocated to a cluster $c \in G(g)$, if flatness metric $\delta_c$ : minimum($\delta_a$) $\forall$ a $\in$ G(g) and *c* is feasible with respect to *p*. If no such *c* is found, *p* is matched to the clusters in the group of clusters G(g+1). This process of matching the incoming partition to groups in G is repeated until a suitable cluster *c* is found to add *p*. If no such cluster is found, a new cluster *c`* is created and *p* is added to *c`*. The algorithm returns the resulting clustering $C_{k+1}$ with *k+1* partitions.

The grouping operation matches incoming partitions to servers that have an available capacity similar to the capacity required by the partition, aiming to result in servers that are fully utilized. The assignment operations chooses the server among the matching clusters that leads to the least flatness metric among the matched group, attempting to achieve a balanced workload on the server. With a worst-case performance of *O*(*N*) where *N* is the number of clusters, Algorithm 3 provides an efficient method to add a partition to an existing clustering.

To *delete* a partition, which is triggered by partition merging, we first remove that partition from the cluster it was assigned to. *ICEPS* then assigns the merged partition to a suitable cluster. An *update* to the access pattern of a partition is considered as a deletion of the partition followed by an insertion of the updated partition.

**Algorithm 4.3:** Update partition plan

**Input**: Partition *p* to be inserted, Clustering $C_k$ of *k* partitions in *N* clusters, integer *numGroups*
**Output**: Clustering $C_{k+1}$ of *N* clusters with *p* assigned to a suitable cluster

$C_{k+1}$ = *null*
cluster *destination* = null
Map<*cluster, integer*> *G* = null
**//*Grouping Operation***
**for each** cluster *c* **in** $C_k$
    integer *group* = (available_capacity(*c*) / max_server_capacity ) * *numGroups*
    *G*.add(*group*, *c*)
**end for**
**//*Assignment Operation***
integer *target_group* = required_capacity(*p*)/max_server_capacity
**while** *target_group* < *numGroups*
    cluster [] *matching_clusters* = *G*.get(*target_group*);
    *destination* = bestFit_cluster **(***matching_clusters***)**
    **if** *destination* != null
        **break**
    *target_group++*
**end while**
**if** *destination* == null
    $C_{k+1}$ **=** add_new_server($C_k$, *p*)
**else**
    $C_{k+1}$ = $C_k$.addPatition_toCluster*(p, destination)*
**return** $C'_{k+1}$

## 4.3 Query Processing

In this section, we show how we use *CEPS* to evaluate circular range query, a popular spatial query in location based applications. Our flat design **increases parallelism** in query processing and adapts a cost-efficient scheme.

The common approach in distributed query processing is employing two types of servers: data and routing [Litwin 1996, Mondal 2004, Mouza 2009, Wang 2010, Ding 2011]. For example, in SD-Rtree (a distributed R-tree) [Mouza 2009], each leaf node corresponds to a data server and each non-leaf node corresponds to a routing server. The problem with this design is that traditional top-down traversal of the tree overloads the servers near root; therefore, it keeps many of the

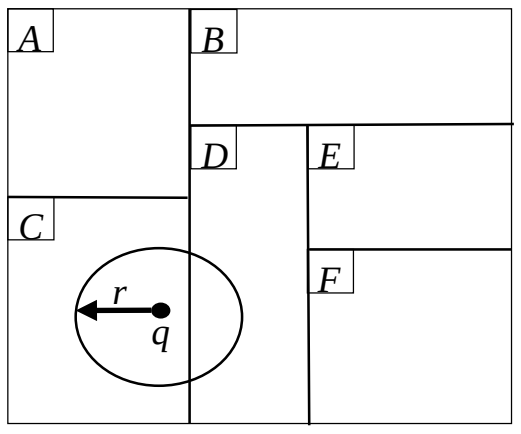

a) Sample query

| IP address | Partitions |
| --- | --- |
| *IP*($S_2$) | *D, E, F* |

b) IP table of $S_1$

Figure 4.3: Range query processing.

servers idle and incurs significant amount of server cost. With our approach, there is no need to allocate extra routing servers due to the flat structure of *CEPS* where each server maintains the coordinates of the partitions, which is significantly smaller than the actual dataset.

More specifically, each server $S_i$ maintains a routing table storing the IP addresses of other servers and list the partitions they store. Let $S_{init}$ represent the server that receives query *q (the location of q),* and $S_q$ represent the server that contains location of *q*. If *q* is contained in $S_{init}$ (i.e., $S_{init}$=$S_q$), the query is processed in $S_{init}$. Otherwise, $S_{init}$ finds $S_q$ by locating the partition that contains *q*, and then forwards *q* to $S_q$ with **one network** message. We use a two-way binary search, first on *x* then on *y* coordinates, to locate $S_q$ on the partition lists. The total cost of this search is ***O*(*log*(*p*))**, where *p* is the number of partitions. Figure 3 illustrates an example of a range query. Let servers $S_1$ and $S_2$ store partitions {*A*, *B*, *C*} and {*D*, *E*, *F*} respectively and $S_{init}$=$S_1$. Upon receiving query *q*, $S_1$ directly locates partitions *C* and *D* which overlap the query region *r* and forwards $q_2$ to $S_2$ as it contains *D*. In this case, the query is processed by two servers in parallel.

We observe that, with *CEPS*, neighboring partitions are stored in different servers to balance the load over time which increase the chance of processing queries in parallel resulting in **lower query response times** especially for larger ranges. Finally, we note that other spatial queries such as *k Nearest Neighbor* (*k*NN) can be implemented as a series of range queries in a distributed fashion [Wang 2010, Akdogan 2014]. In other words, *k*NN query performance depends on the effectiveness of range query; therefore, we do not discuss *k*NN in the rest of the paper.

## 4.4 Experimental Evaluation

In this section, we first present our experimental setup that includes the details about the hardware specifications, metrics, datasets, workloads, and competitor approaches. Subsequently, we present the results and discussions.

### 4.4.1 Experimental Setup and Methodology

We conducted our experiments on Amazon EC2's general-purpose servers. Table 1 (see Section 1) shows the hardware specifications and prices of available servers. The metrics we use in the experiments are as follows: 1) *Server cost* to evaluate the effectiveness of CEPS algorithms, 2) the total amount of n*etwork transfer* to test ICEPS and 3) *Query throughput* (processed requests per second).

#### 4.4.1.1 Dataset & Workload:

We use two different datasets: 1) real-world Gowalla dataset (GD) [Gowalla 2010] and 2) synthetic dataset (SD) which can capture the real-world scenarios where objects cluster around certain location (e.g., cities, point of interests). Gowalla is a location-based social network allowing users to check in and GD includes 6442890 check-ins at 1280969 different locations. Since the number of check-ins is relatively small and it span across almost a two year period, we increase the check-in number 1000 times for every timestamp while keeping other parameters fixed, thus preserving the existing workload distribution over time. Figure 4.4 exemplifies the workload-skew in GD dataset.

With SD dataset, we use Gaussian distribution where the objects are clustered around hotspots (hs) which are initialized at random locations on land coordinates excluding seas and oceans. We vary the number of hotspots from 10 to 1000 where the dataset with 10 hotspots is the most skewed and the one with 1000 hotspots is uniform. The distance from the object to the hotspot follows a Gaussian distribution. In addition, to simulate crowded and less crowded cities, we assign a number s (denoting size) to each hotspot in the range of [1, 100]. Higher s value yields more crowded cities and the largest one will be 100 times larger than the smallest. Furthermore, rather than randomly assigning an s value to a hotspot, we consider the number of internet users [IS 2014] in the city where the hotspot is located. Specifically, s value of a hotspot is calculated based on the normalized value of its corresponding internet user population. Obviously, this approach yields more realistic data distributions since it avoids to generate large clusters where no one lives in or there is no internet user. Finally, SD dataset consists of 40 billion objects in 2D space where each object has an x and y value taking up 4 bytes each.

We generate workloads with cosine and zipf distributions. With cosine, we fix the wave frequency to 1 representing the workload pattern of a day. With zipf, we vary the skew factor (sf) from 0 (uniform) to 3 (skewed). Workloads are associated with the data density. Therefore, as the density increases the magnitude of the workload increases proportionally and we obtain skewed workload at different locations.

#### 4.4.1.2 Competing Approaches:

We evaluate the following approaches in our experiments.

GP: Grid based (spatial) partitioning where each server corresponds to a partition and the workload *is not considered*. The state-of-the-art distributed spatial index, RT-CAN [Wang 2010], adapts this strategy.

GP-R: Grid based partitioning where more *granular* partitions are distributed across the servers randomly. To the best of our knowledge, this strategy is commonly applied in practice.

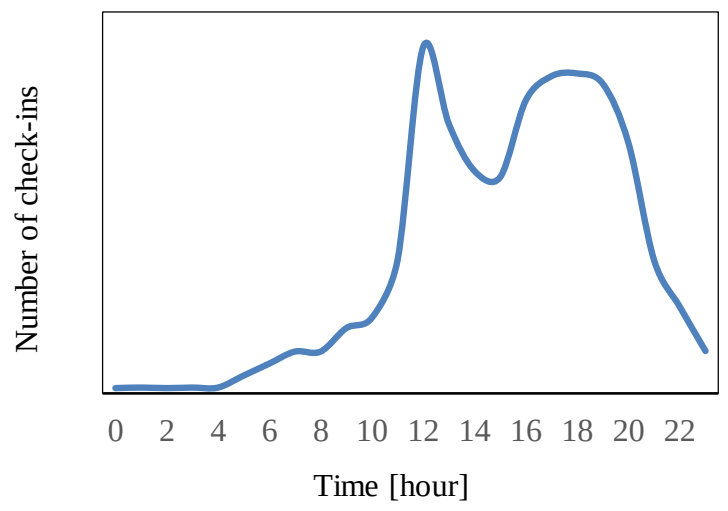

Figure 4.4: Example workload skew. Gowalla check-ins on Monday in GMT timzone.

GP-AAS: Grid based partitioning combined with Amazon's Auto Scaling engine (AAS) which considers *work load* to adjust the amount of resources. This state-of-the-art reactive model is Amazon EC2's primary solution to scale up and down to reduce server cost. AAS requires users to 1) select a monitoring metric and 2) specify their own scaling policy (e.g., when to shut down a server or scaling up to a higher configured server). For our experiments, we monitor *CPUUtilization*, *DiskWriteOps and DiskReadOps* to trigger scaling actions. We also define an effective scaling policy consisting of the following rules. We downgrade to a lowered configuration if the usage is less than 50% for the past 30 minutes. This threshold (50%) is chosen in accordance with EC2 server configurations since a lower profiled server is always half as powerful as the current one (see Table 1 in Section 1). We wait for 30 minutes to avoid workload fluctuations resulting in contradictory scaling up and down requests. If any of the monitoring metrics attains 100% capacity, we can either upgrade to a higher level server or split the data and migrate it to another server. Since the higher level server is twice more costly, in order to save cost we split the data into two, start a new server with the lowest configuration possible which can store the data. This strategy is adapted by distributed hash tables as well that aim to avoid hotspots by sharing data zones [Ratnasamy 2001]. Finally, in order to shut down underutilized servers, we check the last 30 minutes and merge two servers if one of them can handle the workloads of both servers and has enough capacity to store the data.

CEPS-: Our approach where tabu search is disabled.

CEPS+: Our approach where tabu search is enabled.

ICEPS: Incremental CEPS that balancing the load as the access patterns evolve.

We discretize the workload patterns into *5 minute* intervals for our approaches.

### 4.4.2 Experimental Results

#### 4.4.2.1 CEPS:

Given real-world GD dataset, we first investigate how the server cost varies for competing solutions. In this specific experiment, we use Monday workload patterns and set the number of objects at each partition to 5,000 for our approach. For GP, we set the grid capacity to 100,000, which is a relatively small number due to the fact that the size of the actual data is small as well. On the other hand, the workload on each partition is large enough to saturate the computation capacity of the servers. We ensure that each partition is stored on the server with lowest configuration that can handle the workload to reduce the cost and eliminate over-capacity problem. As shown in Figure 4.5, CEPS+ outperforms other approaches and conventional grid-based partitioning significantly increases the server cost due to severe workload skew. GP-AAS is insufficient to adjust itself to fluctuations. We observe that, during morning and afternoon, servers run a little over 50% capacity for several hours which is the worst case for GP-AAS since the underutilized servers cannot be downgraded.

We now vary the number of hotspots on SD dataset, fixed the cluster size ($s$) and investigate the impact of data skew. As the results in Figure 4.6 show, CEPS+ reduces server cost approximately 25% and 40% with respect to GP and GP-AAS. The reason is cosine

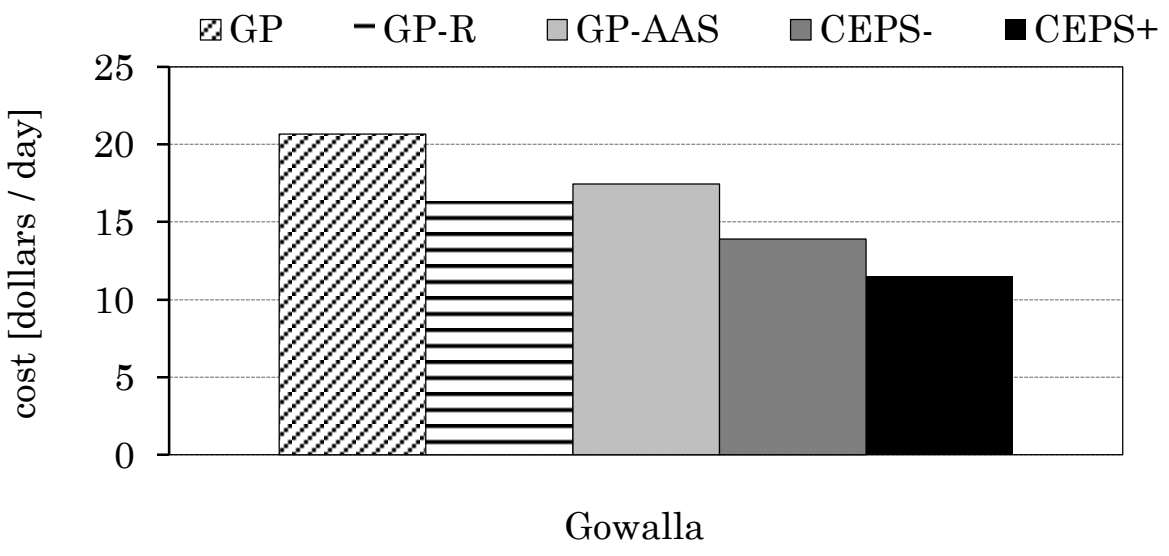

Figure 4.5: Server cost of GD

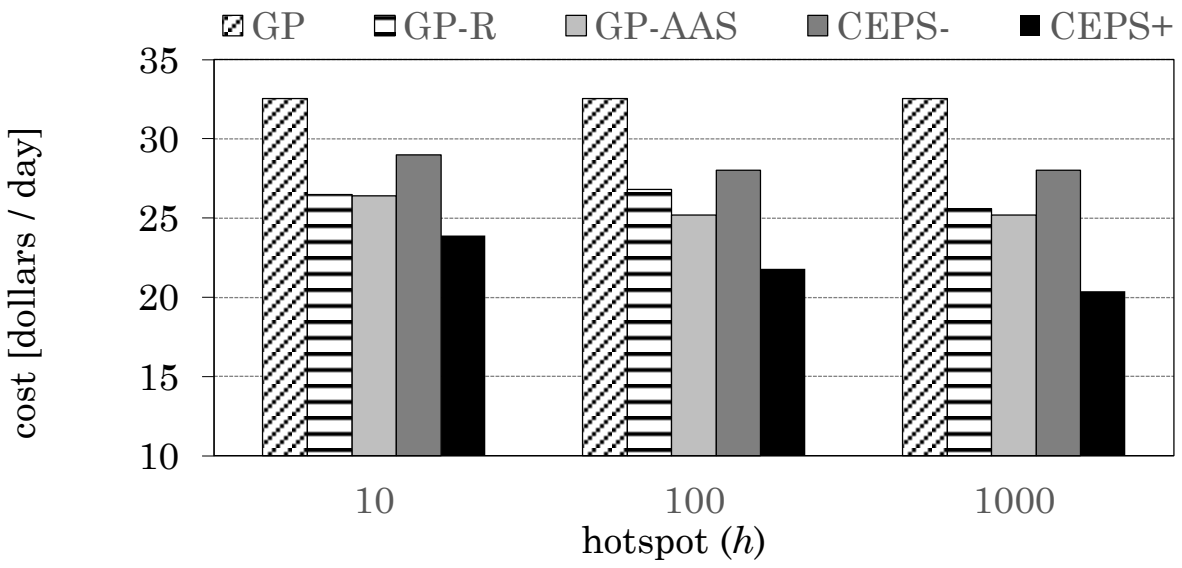

Figure 4.6: Impact of data skew (SD, Cosine, *s*= fixed)

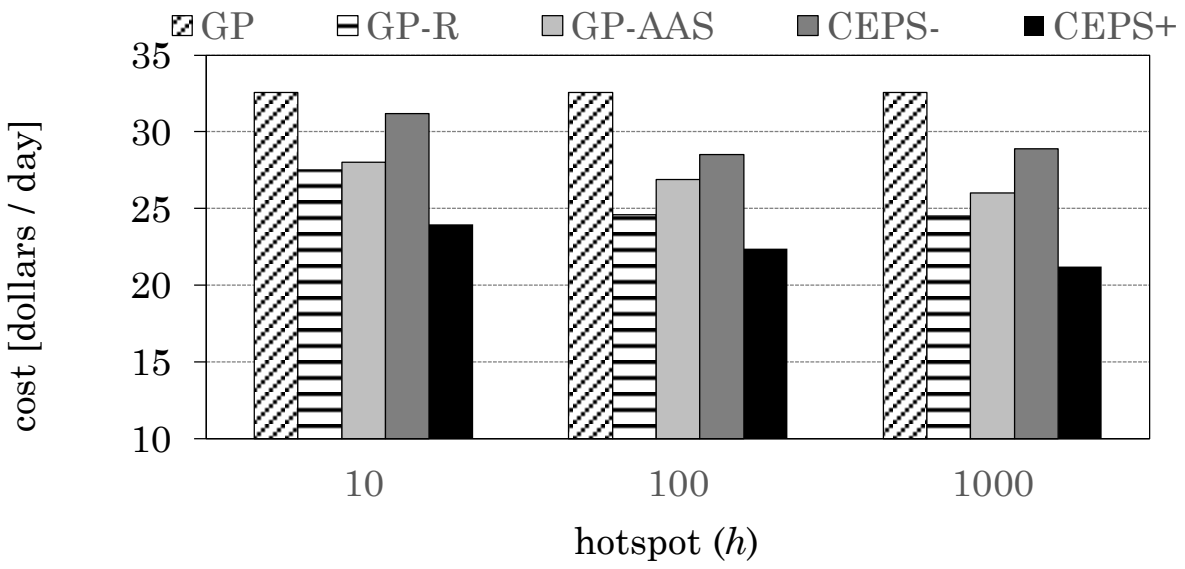

Figure 4.7: Impact of cluster size (SD, Cosine, s=[1,100])

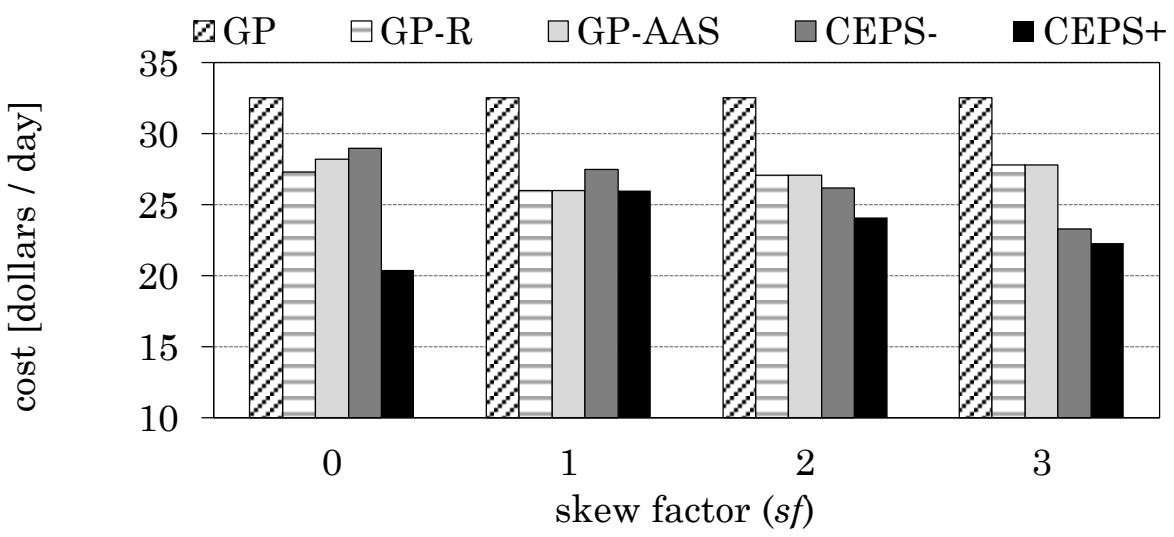

Figure 4.8: Impact of workload skew (SD, Zipf, *hs*=1000)

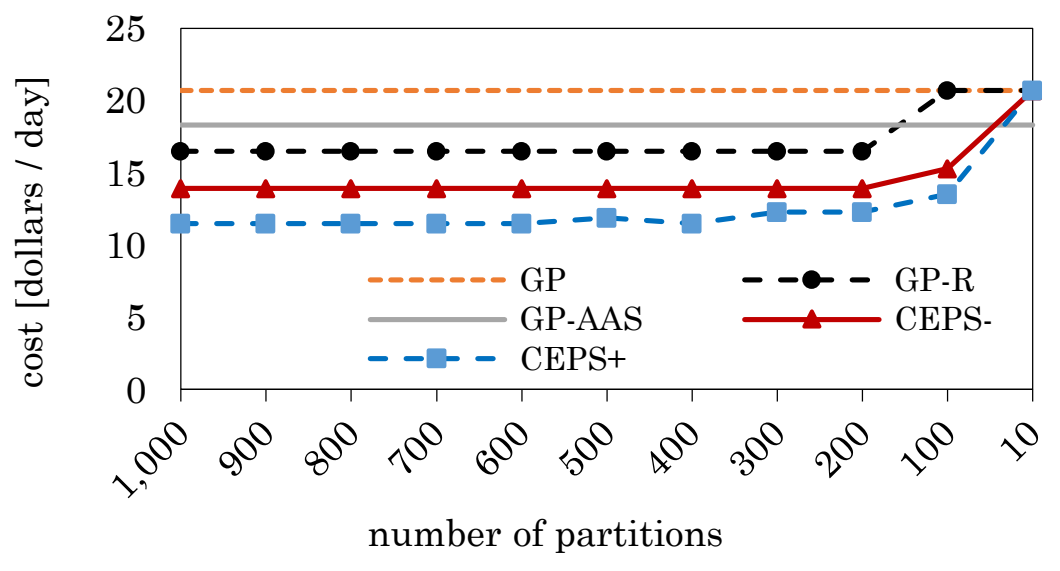

Figure 4.9: Impact of partition number (GD)

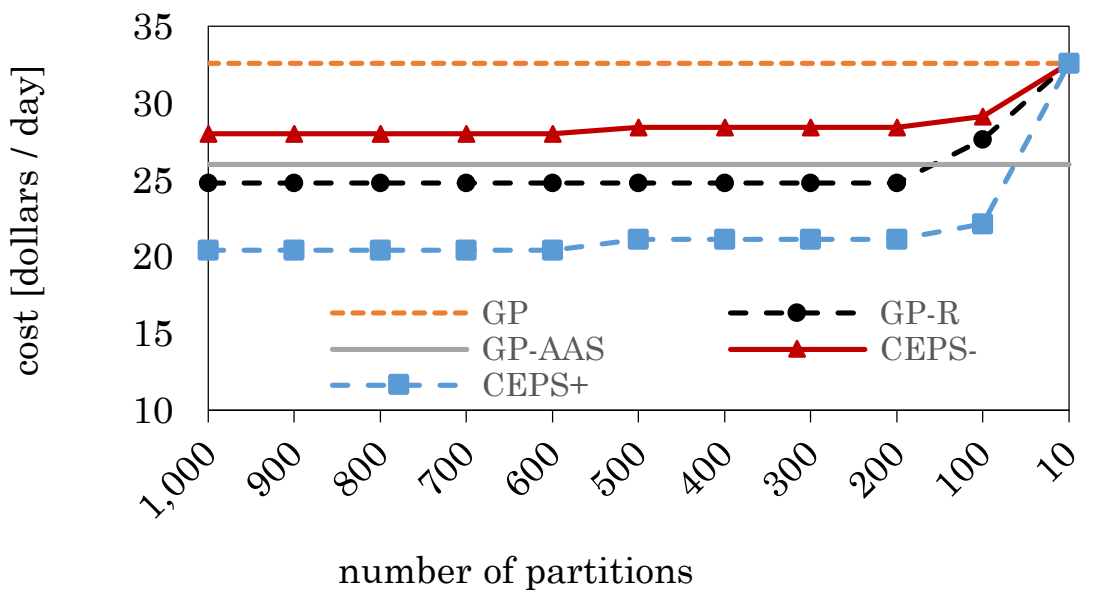

Figure 4.10: Impact of partition number (SD, *hs=1000*, *s*=fixed)

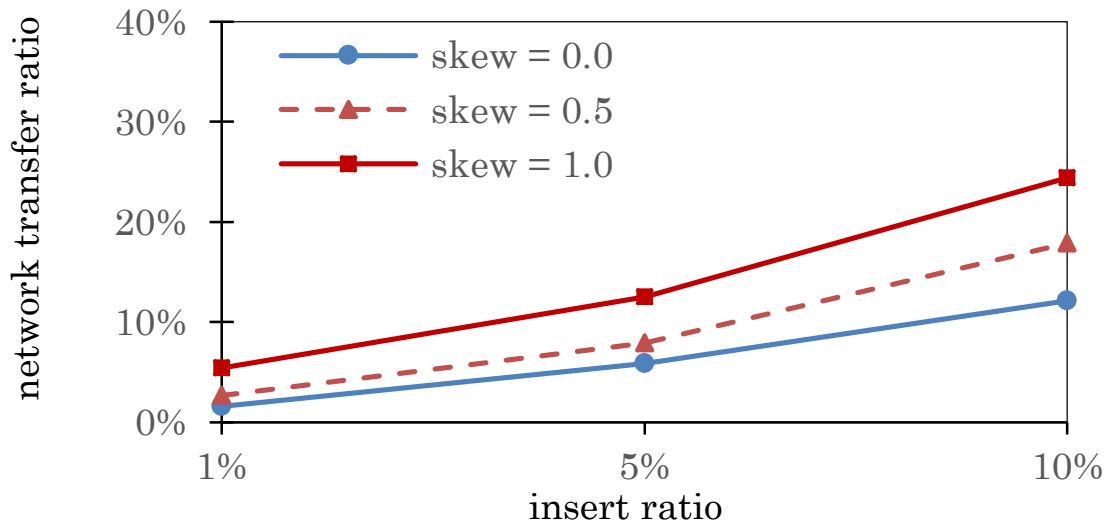

Figure 4.11: Impact of insert ratio and skew factor on ICEPS (GD)

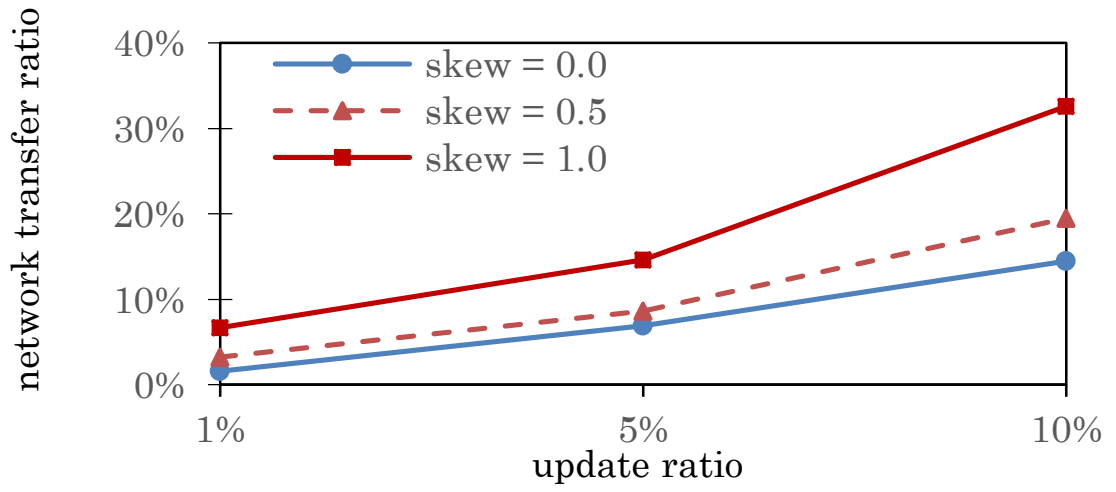

Figure 4.12: Impact of update ratio and skew factor on ICEPS (GD)

workload is the best case scenario for CEPS+ since access patterns are perfectly complementing each other. On the contrary, CEPS- suffers from cosine workload as it directly minimizes the intra-cluster diversity due to these matching patterns. We also observe that, hotspot count does not have a substantial impact on the server cost since the patterns still balance each other.

In the next experiment, we create clusters with various sizes by assigning *s* values in the range of [1, 100]. As Figure 4.7 shows, for small number of hotspots, CEPS+ still outperforms other approaches but the total saving on cost is lower. This is because there are not enough patterns that can completely balance the crowded regions. However, as we increase the number of hotspots, CEPS+ still provides approximately 20-25% cost saving.

Now we experiment with a different workload (zipf), fix the cluster size, set the number of hotspots to 1000 and vary the skew factor from 0 (constant) to 3 (skew). Skewed workload indicates that the data is accessed for a short period of time throughout a day. As Figure 4.8 illustrates, when the workload is constant, CEPS+ outperforms other approaches since all the patterns complement each other. Similar to cosine workload CEPS- suffers from maximized intra-cluster diversity. We also observe that with low-skew (*sf*=1), the gain obtained by adapting CEPS+ slightly decreases with respect to GP-AAS.

In the next experiment, we investigate the impact of the number of partitions (p) on server cost by varying the number of objects at each partition on GD dataset. As shown in Figure 4.9, the total cost slightly increases similar to a step function as the number of partitions decreases and when the p is very small (e.g., 10) the total costs are the same. This is because fine-granular partitions (access patterns) improve the quality of AHR clustering. In addition, CEPS- is slightly better than GP-R. We also observe that after a few hundred partitions the result does not change substantially. One may argue that small partitions degrades query performance; however, with spatial data, small partitions are sufficient to group co-accessed data objects together that reduces the local disk I/O at each server during query processing. For example, in real-world applications, users only query a limited region around them (e.g., show me the Starbucks shops within 15 miles). We repeat the same experiment on SD dataset with cosine workload where h=1000 and s is fixed. As Figure 4.10 illustrates, as long as we generate more than 100 partitions, we can reduce the server cost up to 40%.

#### 4.4.2.2 Incremental CEPS:

We now evaluate the performance of ICEPS against insert and update requests. Specifically, we vary the number of requests, the distribution that requests follow and report the network transfer ratio (total amount of data transfer / the size of entire dataset). We use this metric since we aim to minimize network transfer while maintaining our objective. Recall that re-executing CEPS from scratch whenever a partition is modified, which is frequent due to granular partitions, would require redistributing the entire dataset where networks transfer ratio is 100%.

In this experiment, we insert new data objects following zipf distribution on GD dataset. When skewed factor is set to 0, the new data objects follow a uniform distribution. When skewed factor is set to 1, about 80% inserts focus on 20% of the partitions. In addition, we vary the amount of new objects. Specifically, we consider three separate insertion ratios: small (1%), medium (5%) and large (10%) where the percentage indicates the ratio of new data objects to the existing dataset. Since the data is not associated with any workload, we assign every new object the average workload of the partition that it is inserted in. As the results in Figure 4.11 show, as the insertions follow skew-distribution, the total amount of network I/O increases compared to the uniform distribution. Because skew-distribution causes partition splits which triggers ICEPS to maintain the objective function. However, the total amount of network transfer is still significantly lower than the actual data size even with the skew-distribution.

Figure 4.12 illustrates the results with the same experimental setup for update requests. With update, we generate a random number between 0 to 100 miles and a direction (e.g., north, south) chosen randomly and change the current location of the objects based on these two parameters. We observe that update incurs more network I/O. Because, an update request consists of consecutive delete and insert operations, thus the number of partitions which are affected by an update is higher.

#### 4.4.2.3 Query Processing:

We now measure query throughput to evaluate the performance of range queries while varying query range. We define range as distance in mile with respect to a query point, and as proposed in [Wang 2010] we consider three separate selectivity ranges: small (5 miles), medium (25 miles) and large (50 miles). As discussed earlier, CEPS+ generates smaller partitions and clusters them based on workloads and GP only considers the spatial proximity among the data objects. To evaluate the impact of both approaches on query throughput, we fix the number of servers, partition GD dataset using only GD and CEPS+ where the number of objects is set to 100,000 for GP and 10,000 for CEPS+. We randomly generate the query points and send each query to servers for both approaches in a round-robin fashion. As for the query routing, we apply the same technique and forward the queries to the corresponding servers with one network message.

As the results in Figure 4.13 show, the query throughput decreases as the range enlarges for GP approach. This is because we need to retrieve more objects which degrades the performance. On the contrary, with CEPS+, query throughput slightly decreases as the query range increases. The reason is that spatially close partitions are more likely to be stored in different servers. Therefore, as the query range gets larger, more nodes are involved in query processing and we take advantage of parallelism which verifies the effectiveness of CEPS+. We also observe that, with the small range (5 miles), both approaches have similar performances since the query coordination cost of parallel processing in CEPS+ does not pay off for queries with small selectivity. As shown in Figure 4.14, we observe the same pattern for SD dataset with cosine workload where throughput slightly decreases for CEPS+ and substantially diminishes for GP.

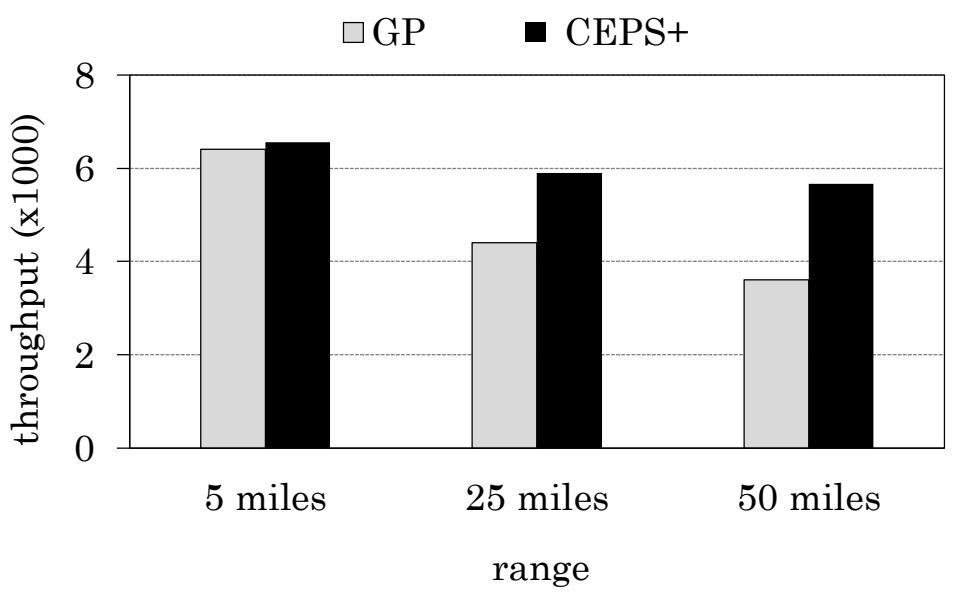

Figure 4.13: Impact of radius on range query (GD)

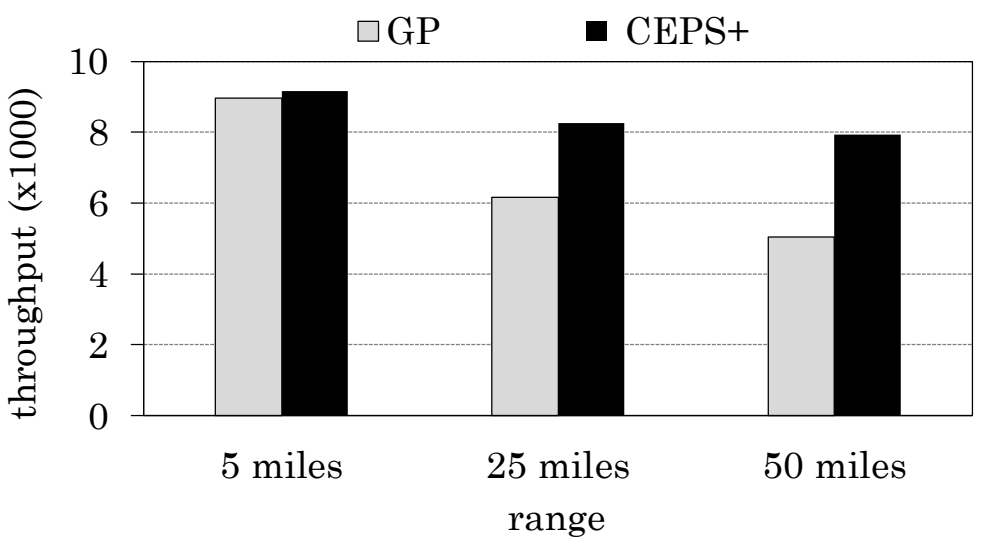

Figure 4.14. Impact of radius on range query (SD, *s*=fixed, *hs*=1000)

## 4.5 Discussion

Finally, we would like to emphasize two key points that augment the applicability of our approach.

**Replication:** There are two basic types of spatial replication techniques: 1) border-only approach where only the data objects at the partition borders are replicated [Akdogan 2014] and 2) multiple-copy strategy where the partitions are copied entirely and stored on multiple servers for data availability in case of a failure. We can easily adapt both replication techniques into CEPS without increasing the time complexity. For the border-only strategy, simply the data partitioning phase needs to be modified in a way where each partition covers a certain portion of neighboring partitions. For the second technique, the input to AHR clustering needs to be the original partitions and their replicas. We also need to ensure that two replicas of the same partition are not stored in the same server for availability purposes. This constraint can easily be handled through tabu list.

**Multi-dimensional datasets:** Even though we use datasets in 2D spaces for experiments, we argue that our approach can be applied to higher dimensional spaces by slightly modifying the partitioning technique which can partition n-dimensional data.

## 4.6 Summary of Chapter

We proposed a new cost-oriented partitioning method for spatial datasets that maximizes server utilization and saves server cost. Our problem is motivated by several pragmatic considerations such as time-based pricing models of the cloud providers, limitations in storage services that do not support advanced queries required by location-based services, complex configuration of scaling engines and coarse-grained servers. Our experiments show that CEPS reduces cost by up to 40%, enhances parallelism in query processing and increases query throughput as compared to grid-based partitioning.

We believe that our approach can be adapted by the global location-based services that are becoming increasingly popular with the recent advances in wearable technologies, smart phones, location-aware Internet browsers and offline geo-tagging tools.

# Chapter 5: Conclusions

In this thesis, we focused on scalable, efficient and cost-effective management of spatial data on cloud. Towards this end, we first focused on handling dynamic spatial datasets which are commonly generated by mobile applications such as ride-sharing services, mobile social networking, etc. These applications need to handle tremendous number of spatial objects that continuously move and execute spatial queries to explore their surroundings. To manage such datasets effectively, we first introduced a novel distributed throwaway index, *D-ToSS*, that partitions and indexes a dynamic spatial datasets using Voronoi diagram which we also use for query processing. *D-ToSS* not only scales out to multiple servers but also scales up since it fully exploits the multi-core CPUs available on each server. *D-ToSS* yields a very high query throughput, scales **near-linearly** with the number of nodes in the cluster and the total execution time remains almost constant as the data size increases. Our experiments show that if **2+%** of the objects issue update to the index, it is faster to rebuild *D-ToSS* than update RT-CAN, the state-of-the-art distributed incremental index structure. In addition to scalability, our experiments show that we can generate an index for approximately 67 million objects just within **a few seconds** using a small cluster of servers on Amazon's EC2 infrastructure, which verifies the efficiency of D-ToSS.

In addition to scalability and efficiency, we also aimed to maximize the server utilization that can support the same workload with less number of servers. Server utilization is a crucial point while using cloud computing because users are charged based on the total amount of time they reserve each server, with no consideration of utilization. Therefore, it is essential for users to fully utilize all servers to reduce the cost. Clearly, one key factor that impacts server utilization is the partitioning method especially in data-driven traffic/transportation services. This is because if the data partitions are not accessed, the servers storing them remain idle but the user is still charged.

Whereas, conventional spatial partitioning techniques only consider the data distribution cluster spatially close objects to minimize disk I/O. On the contrary, the objective is different for cloud given the current pricing models. To reduce the server cost, we devised a novel cost-efficient partitioning method for spatial data where an increase in the servers' utilizations yields less number of servers to support the same workload, thus saving cost. Our approach considers both 1) spatial proximity among the data objects and 2) access patterns (workload) associated with the objects (e.g., how many times an object is accessed at a particular time). We conducted extensive experiments on EC2 using real-world and synthetic datasets and observed that we can reduce the server cost up to 40%.